\documentclass[smallextended]{svjour3} 

\usepackage{graphicx}
\usepackage{amssymb}
\usepackage{amsmath}
\usepackage[latin1]{inputenc}

\usepackage[T1]{fontenc}
\usepackage{ae,aecompl,url}
\usepackage{multicol}
\usepackage{cuted,mathtools}

\title{Tilting Styx and Nix but not Uranus with a Spin-Precession-Mean-motion resonance}

\author % [Quillen et al.]
{
Alice C. Quillen$^{1}$,  
Yuan-Yuan Chen$^{1,2}$,     Beno\^it Noyelles$^3$, \&  
Santiago Loane$^1$
}
\institute{ \at
$^1$Department of Physics and Astronomy, University of Rochester, Rochester, NY 14627 USA \\
\email{alice.quillen@rochester.edu} 
\and \at
$^2$Key Laboratory of Planetary Sciences, Purple Mountain Observatory, Chinese Academy of Sciences, Nanjing 210008, China
\and \at
$^3$Department of Mathematics and the Namur Centre for Complex Systems (naXys), \\
 University of Namur, 8 Rempart de la Vierge, Namur B-5000 Belgium
}

\titlerunning{Tilting Styx and Nix}
\authorrunning{Quillen et al.}

\begin{document}
\maketitle

\begin{abstract}
A Hamiltonian model is constructed for the spin axis of a  planet 
perturbed by a nearby planet with both planets in orbit about star.
 We expand the planet-planet gravitational potential perturbation to
first order in orbital inclinations and eccentricities, finding terms describing  spin resonances
involving the spin precession rate and the two planetary mean motions.
Convergent planetary migration allows the spinning planet to be captured
into spin resonance.  With initial obliquity  near
zero, the spin resonance can lift the planet's obliquity to near 90 or 180 degrees
depending upon whether the spin resonance is 
first or zero-th order in inclination. 
Past capture of Uranus into such a spin resonance could give an alternative
 non-collisional scenario accounting for Uranus's  high obliquity.
However we find
that the time spent in spin resonance must be so long that this scenario cannot be
 responsible for Uranus's high obliquity.
Our model can be used to study spin resonance in satellite systems.
Our Hamiltonian model explains how
Styx and Nix can be tilted to high obliquity via outward migration of Charon, 
a phenomenon previously seen in numerical simulations.
 
\end{abstract}


%\section{Todo:}

\section{Introduction}

Resonances involving planet or satellite spin can cause chaotic tumbling, 
prevent a body from tidally despinning or affect obliquity. 
Mercury was captured into a spin-orbit resonance, with spin rate
 a half integer multiple of its orbital mean motion, 
 (e.g., \cite{goldreich66,noyelles14}) whereas Hyperion is  
chaotically tumbling due to spin-orbit resonance overlap \cite{wisdom84}.
Secular spin-orbit resonances occur when the period of precession of the spin
axis of a planet  is commensurate  with  (an integer multiple of) one of the periods of 
secular orbit variation  \cite{ward74}.
 Chaotic obliquity variations in Mars  is attributed to secular spin resonances
\cite{ward73,ward74,laskar93,touma93}.
Capture into the secular spin resonance connected to the vertical secular eigenfrequency 
associated with Neptune
may have tilted Saturn's obliquity to its current value of 26.7$^\circ$ \cite{ward04,hamilton04}.
For an satellite in orbit about a binary such as  Pluto and Charon,  spin-binary resonance involves 
 a commensurability between the binary mean motion, the orbital mean
motion and the satellite spin rate \cite{correia15}.  
So far we have mentioned three-types of spin resonance: spin-orbit resonance (Mercury, Hyperion), 
spin secular resonance (Mars, Saturn) and spin-binary resonance.

Our numerical study of obliquity evolution of Pluto and Charon's minor satellites,
showed another type of spin resonance \cite{quillen17}.
We found that a commensurability involving a mean motion resonance between Charon
and a minor satellite and  the satellite's spin precession rate could influence its obliquity. % \cite{quillen17}.
For satellite Styx, near a 3:1 mean motion resonance with Charon, we saw obliquity variations
when the angles 
\begin{align}
\phi_{s1} &= 3 \lambda_{Styx} - \lambda_{Charon} - \phi_{Styx} - \Omega_{Styx}  \nonumber \\
\phi_{s2} &= 3 \lambda_{Styx} - \lambda_{Charon} - 2\phi_{Styx}  \label{eqn:phi_s}
\end{align}
were librating about constant values rather than circulating.  
Here $\lambda_{Styx}$ and $\lambda_{Charon}$ are the  mean longitudes of Styx and Charon
and $\Omega_{Styx}$ is the longitude of the ascending node of Styx.  Orbital elements
are measured with respect to Pluto or the center of mass of the Pluto/Charon binary, 
and its satellite orbital plane, not the Sun and the ecliptic.   
The precession angle $\phi_{Styx}$ describes the orientation of Styx's spin axis with
$\dot \phi_{Styx}<0$ as Styx's spin axis precesses about the orbit normal.

The New Horizons Mission found that Pluto and Charon's minor satellites, 
Styx, Nix, Kerberos and Hydra have 
not tidally spun down to near synchronous rotation and that all of them
have high obliquities near $90^\circ$ \cite{weaver16}.
\cite{quillen17} suggested that the minor satellite current obliquities need not be primordial. 
A spin resonance  involving a mean motion resonance between Charon
and a minor satellite and  the satellite's spin precession rate, when drifting due to an outwards migrating Charon, can lift the obliquities of the minor satellites, 
accounting for their high and near $90^\circ$ obliquities
discovered by the New Horizons Mission.

As we were lacking a model for this type of spin resonance strength, we were unable
to assess its strength or even identify which type of resonant angle was likely to be most important
for each of Pluto and Charon's minor satellites.
We address this issue here with the development of a Hamiltonian model for this spin resonance
in section \ref{sec:spin_ev}. In section  \ref{sec:spin_toy} we explore resonance
capture by allowing the resonance in our Hamiltonian
model to drift.   In section \ref{sec:pc} we apply our model to Pluto and
Charon's minor satellites.  

Uranus  has a high obliquity, of 98$^\circ$.
Could  a similar spin resonance  have tilted Uranus during a previous time when Uranus was
in or near a mean motion resonance with another giant planet?    Using the Hamiltonian model
of sections \ref{sec:spin_ev} and \ref{sec:spin_toy} we answer this question in section \ref{sec:uranus}.
To aid the reader, a list of symbols is included in Table \ref{tab:tab}.

\begin{table}
\vbox to150mm{\vfil
\caption{\large  List of Symbols \label{tab:tab}}
\begin{tabular}{@{}lllllll}
\hline
$a$ & semi-major axis \\
$e$ & orbital eccentricity \\
$I$ & orbital inclination \\
$\Omega$ & longitude of the ascending node \\
$\omega $ & argument of pericenter \\
$M$  & mean anomaly \\
$\varpi = \omega + \Omega $ &  longitude of pericenter \\
$ \lambda = M + \varpi$ &  mean longitude \\
$M_*$ & mass of central star \\
$n$ & mean motion \\
$\hat {\bf n}$ &  orbit normal unit vector \\
$\hat {\bf s}$ & spin direction unit vector \\
$C,A$ & moments of inertia of oblate planet \\
$w$ &  spin angular rotation rate of planet \\
$\alpha_s$ & spin precession rate \\
$\alpha$ & ratio of semi-major axes \\
${\bf T}$ & torque vector \\
$r$ & orbital radius \\
$\phi$ & spin precession angle \\
$\theta$ &    obliquity angle \\
$s\approx I/2$  & used in low inclination expansions \\ % s = sin (I/2)
$t$  & time \\
$ \tau$ & normalized time \\
$R$  & spinning planet's equatorial radius \\
$J_2$ & second zonal gravity harmonic for the spinning planet \\
$q_s$ & normalized quadrupole coefficient of satellite system \\
$l_s$ & normalized angular momentum of satellite system \\
$\lambda_C$ & normalized moment of inertia about principal axis \\
$p$ & canonical momentum variable, a function of obliquity  \\
$\Delta$ & distance between perturber and spinning body \\
$\psi, \Psi$ & angles used in expansion of disturbing function \\
$b_{s}^{(j)}(\alpha)$ & Laplace coefficient \\
$j$  & resonance index \\
$\epsilon$ & resonance strength in a Hamiltonian model \\
$\nu$   & distance to resonance in a Hamiltonian model \\
$c_0^j, c_s^j, c_{s'}^j$  & coefficients used to compute spin resonance strengths  \\
$c_{e1}^j, c_{e'1}^j$  & ''  \\
$ c_{e3}^j,c_{e'3}^j$ & '' \\
$\beta$ & '' \\
\hline
\end{tabular}
{\\ %Notes
}}
\end{table}

\section{Spin evolution}
\label{sec:spin_ev}

A spinning oblate planet in orbit about a central mass $M_*$ that has spin axis, $\hat{\bf s}$,
tilted with respect to the orbit plane, precesses.
We refer to the spinning object as a planet in orbit about a star, however we keep in mind 
that we can also consider a spinning satellite in orbit about a planet, asteroid or Kuiper belt object.
We assume that the planet is rapidly spinning about
its principal inertial axis and this is known as the {\it gyroscopic approximation}.
The planet's moments of inertia are $A,C$ with $A>C$ and
the planet's spin angular momentum is ${\bf L}_s = w C \hat {\bf s}$
 where $w$ is the spin angular rotation rate and $\hat {\bf s}$ is a unit vector.
The planet's spin axis satisfies
\begin{equation}
\frac{d \hat {\bf s}}{dt} = \alpha_s (\hat {\bf s} \cdot \hat {\bf n} ) (\hat {\bf s} \times \hat {\bf n} ) 
\label{eqn:dsdt}
\end{equation}
 \cite{colombo66},  with time derivatives taken with respect to the inertial frame. 
 Here $\hat {\bf n}$ is a unit vector perpendicular to the orbit plane, aligned with the orbital
angular momentum vector. 
The precession rate
\begin{equation}
\alpha_s = \frac{3}{2} \frac{(C-A)}{Cw} \frac{n^2}{(1-e^2)^\frac{3}{2}},  \label{eqn:alpha}
\end{equation}
where the orbital mean motion  is $n$
and the orbital eccentricity is $e$.  

MacCullagh's formula gives the 
 instantaneous torque on an oblate planet due to point mass $M_*$
 \begin{equation}
 {\bf T} = 3(C-A)\frac{GM_*}{r^3}  
( \hat {\bf r} \cdot \hat {\bf s} ) ( \hat {\bf r} \times \hat {\bf s} ) \label{eqn:mac}
\end{equation}
where ${\bf r} = r \hat {\bf r}$ is the vector between the planet's center of mass and $M_*$.
Equation \ref{eqn:dsdt} can be derived using MacCullagh's formula
for the instantaneous torque by 
averaging over the orbit or computing $\langle {\bf T}\rangle = \frac{1}{P} \int {\bf T} dt$ where
the orbital period is $P=2\pi/n$, and assuming that
the planet  remains spinning nearly about its principal axis \cite{colombo66}.
Thus equation \ref{eqn:dsdt} is consistent with
\begin{equation}
\frac{d\hat {\bf s}}{dt} = \frac{1}{Cw} \langle {\bf T} \rangle .
\end{equation}

Due to secular perturbations arising from other planets, 
the orbit normal $\hat{\bf n}$ is a function of time 
(e.g., \cite{colombo66,ward75}).
A time dependent  equation \ref{eqn:dsdt} has been used 
to study tidal evolution into Cassini states 
\cite{colombo66,ward75} and obliquity evolution of Mars \cite{ward73,ward79,bills90}
and Saturn \cite{ward04}.  Phenomena discovered and explored
include capture into spin-secular resonance states (Saturn; \cite{ward04,hamilton04}) 
and chaotic obliquity evolution (Mars; \cite{ward73,touma93,laskar93}).

\subsection{A Hamiltonian model for spin about a principal axis}

Using angular spherical coordinates $\phi,\theta$  in an inertial reference frame 
 to specify the spin axis
\begin{equation}
\hat {\bf s} = (\cos \phi \sin\theta, \sin\phi \sin \theta, \cos \theta), \label{eqn:s}
\end{equation} 
equation \ref{eqn:dsdt} can be written as a Hamiltonian dynamical 
system with canonical momentum $p$, conjugate to the precession angle $\phi$
\begin{equation}
p = (1- \cos \theta) .
\end{equation}
The  Hamiltonian  
\begin{equation}
H_{ave}(p,\phi) = \frac{\alpha_s}{2} (\hat {\bf s} \cdot \hat {\bf n})^2 \label{eqn:ham_n}
\end{equation}
with $\hat {\bf s}$ a function of $p, \phi$,
(similar to that used by \cite{goldreich69} or \cite{ward91}).
Hamilton's equations are
\begin{eqnarray}
\dot p &=& -\frac{\partial H_{ave}}{\partial \phi} \\
\dot \phi &=& \frac{\partial H_{ave}}{\partial p}
\end{eqnarray}
and are equivalent to the equations of motion for the spin axis in 
equation \ref{eqn:dsdt}.
%Peale 1974 equation 45, 46 and referring to older work by Goldreich and Toomre.

With orbit normal $\hat {\bf n}$ in the $z$ direction, The angle $\phi \in [0,2\pi]$ describes spin precession.  
$\theta \in [0,\pi] $ is the planet's obliquity.  The canonical momentum $p \in [0,2]$
with $p=0$ at $\theta=0$.  With the addition of a third angle describing body orientation
about the spin axis, $\phi,\theta$ are Euler angles.

Equation \ref{eqn:dsdt} for $d \hat {\bf s}/dt $ resembles equation \ref{eqn:mac} for ${\bf T}$
but with ${\bf r}$ replacing $\hat {\bf n}$.
Because position vector $\bf r $ is independent of the spin orientation angles $\theta$ and $ \phi$,
the instantaneous spin vector (prior to averaging over the orbit) 
 can also be described with a Hamiltonian system with  
\begin{equation}
H(p,\phi) = -\frac{3}{2} \frac{(C-A)}{Cw} \frac{GM}{r^3}  (\hat {\bf s} \cdot  \hat{\bf r})^2,
\label{eqn:ham2}
\end{equation}
again with $\hat {\bf s}$ a function of $p, \phi$.
Hamilton's equations are equations of motion equivalent to 
\begin{equation}
\frac{d \hat {\bf s}}{dt} = \frac{1}{Cw} {\bf T}.
\end{equation}
When averaged over the orbit period this equation  of motion for $\hat {\bf s}$
is consistent with equation \ref{eqn:dsdt}.
The Hamiltonian in equation \ref{eqn:ham2} can be averaged by
writing  $r$ and $\hat{\bf r}$ in terms of
 the mean anomaly and mean longitude and taking the
 average over  these angles, yielding equation \ref{eqn:ham_n}.    
The gyroscopic approximation should be a good one as long
as the orbital period is much larger than the spin rotation period.
For rigorous averaging calculations see \cite{boue06}.

\subsection{Precessional Constant with Satellites}

An spinning oblate planet locks its satellites 
 to its equator plane so that the system precesses as a unit \cite{goldreich65}.
The precession rate in equation \ref{eqn:alpha} can be modified to take
into account the satellites with
\begin{equation}
\alpha_s = \frac{3}{2} \frac{n^2}{w} \frac{J_2 +q_s }{\lambda_C + l_s},  \label{eqn:alpha_mod}
\end{equation}
\cite{ward75,french93,ward04}
but neglecting the orbital eccentricity.
Here $J_2$ is the coefficient of the second zonal gravity harmonic (from the quadrupole moment) 
of the planet's gravitational potential field
and $\lambda_C = C/m R^2$ is the planet's moment of inertia about its principal axis normalized
to the product of planet mass and the square of the planet's equatorial radius.
The parameter 
\begin{equation}
l_s \equiv \sum_j \frac{m_j}{m} \left( \frac{a_j}{R}\right)^2  \frac{n_j}{w} \label{eqn:ls}
\end{equation}
is the angular momentum of the satellite system normalized to $mR^2 w$ 
where $m_j, a_j,n_j$ are the masses, semi-major axes (for the orbit about
the planet) and mean motions of each satellite.
The parameter
\begin{equation}
q_s \equiv \frac{1}{2} \sum_j  \frac{m_j}{m} \left( \frac{a_j}{R}\right)^2  \frac{\sin(\theta - I_j)}{\sin\theta}
\label{eqn:qs}
\end{equation}
is the effective quadrupole coefficient of the satellite system with $q_s/J_2$ 
being the ratio of the solar torque on the satellites to that directly exerted on the planet.
Here $\theta$ is the planet's obliquity and $I_j$ the inclination of the $j$-th satellite
with respect to the planet's equatorial plane.
Without satellites $q_s = l_s =0$ and $J_2 = (C-A)/(mR^2)$ so that  equation 
\ref{eqn:alpha_mod} reduces to equation \ref{eqn:alpha} at zero eccentricity.
 
\subsection{A Perturbed Hamiltonian Model}

We consider a spinning planet in orbit about a star 
at zero orbital inclination. Henceforth we take orbital normal $\hat {\bf n} = \hat {\bf z}$.
When averaged over the orbit period and over the longitude of the ascending node
the Hamiltonian describing the planet's spin (equation \ref{eqn:ham_n})
\begin{equation}
H_0(p,\phi) = \frac{\alpha_s}{2} (p-1)^2  \label{eqn:ham0}
\end{equation}
and giving spin precession rate
\begin{equation}
\dot \phi = - \alpha_s \cos \theta
\end{equation}
with $\alpha_s$ as given  in equation \ref{eqn:alpha_mod}.

We consider a Hamiltonian model that includes a perturbation to $H_0$, in the form
\begin{equation}
H(p,\phi,t) = H_0(p) + H_1(p,\phi,t) \label{eqn:hamfirst}
\end{equation}
where $H_1$ is a time dependent perturbation.

MacCullagh's formula gives the torque on our spinning planet due to a
perturbing planet with mass $m_p$.  The radial vector between
the two planets is ${\bf r} - {\bf r}_p$ where ${\bf r}$ is the radial vector to the spinning planet 
(with respect to the central star) and ${\bf r}_p$
the radial vector to the perturbing planet. 
The torque on the spinning planet is dependent upon the radial vector between
the two planets 
 \begin{equation}
 {\bf T} = 3(C-A)\frac{Gm_p}{|{\bf r} - {\bf r_p}|^5}  
\left( ({\bf r} - {\bf r}_p) \cdot \hat {\bf s} \right) \left( ({\bf r} - {\bf r}_p) \times \hat {\bf s} \right) .\label{eqn:mac2}
\end{equation}
 The perturbing planet is treated as a point mass.
The associated Hamiltonian perturbation term (arising from ${\bf T}$) is
\begin{align}
H_1(p,\phi,t) =- \frac{3(C-A)}{Cw} \frac{Gm_p}{|{\bf r} - {\bf r_p}|^5}  \frac{ \left( ({\bf r} - {\bf r}_p) \cdot \hat {\bf s} \right)^2}{2} \label{eqn:H1}
\end{align}
and it is a time dependent perturbation as ${\bf r}, {\bf r}_p$ vary.
$H_1 \ll H_0$ because the mass of the perturbing planet is much
less than the mass of a star; $m_p \ll M_*$.

We describe the orbits in terms of orbital elements $a,e,I,\Omega,M$ which are semi-major axis,
eccentricity, inclination, longitude of the ascending node, argument of pericenter and 
mean anomaly, respectively.  
We also use the mean longitude $\lambda = \Omega + \omega + M$
and the longitude of pericenter $\varpi = \Omega + \omega$.
Our spinning  and perturbing planets orbit a star with mass $M_*$.

 Above  ${\bf r}$ and $ {\bf r}_p$ refer
to positions of spinning and perturbing planets.  
We now depart  from this notation, using 
 ${\bf r}$ and ${\bf r}'$ to refer to radial vectors from the star of inner and outer orbiting masses.
Orbital elements for the object with the {\it larger} semi-major axis will be referred to
with a prime $(a',e',I',\Omega',M', \lambda', \varpi')$ and
those with the {\it smaller} semi-major axis without a prime.
The ratio of semi-major axes $\alpha \equiv a/a'$.
With radial vectors ${\bf r}$ and ${\bf r}'$ for inner and outer orbiting mass, 
equation \ref{eqn:H1} for the perturbation  becomes
\begin{align}
H_1(p,\phi,t) = -3\frac{(C-A)}{Cw} n^2 \frac{m_p}{M_*} 
\frac{a_s^3} {|{\bf r} - {\bf r}'|^5}  \frac{ \left( ({\bf r} - {\bf r}') \cdot \hat {\bf s} \right)^2}{2} \label{eqn:H1p}
\end{align}
where $a_s$ is the semi-major axis of the spinning body,  
\begin{align}
a_s \equiv  \left\{ 
\begin{array}{lr} 
a'  & \text{for external spinning body} \\
a  & \text{for internal spinning body} .
\end{array} \right. 
\end{align}
When the spinning body is external, we mean that it is perturbed by
the mass $m_p$ that has orbit interior to the spinning body.
 %(equal to $a'$ if it has the outermost orbit, otherwise equal to $a$).
 
Taking into account a satellite system around the spinning planet
$$\frac{(C-A)}{C} \to \frac{(J_2+q_s)}{(\lambda_C+ l_s)}$$
 (comparing equation \ref{eqn:alpha}
with \ref{eqn:alpha_mod}) and defining
\begin{equation}
\Delta \equiv |{\bf r} - {\bf r}' |,
\end{equation}
we can write equation \ref{eqn:H1p} as
\begin{align}
H_1(p,\phi,t) =   -\alpha_s \frac{m_p}{M_*} \left(\frac{a_s}{a'}\right)^3 \frac{a'^3}{\Delta^5 }
 \left( ({\bf r} - {\bf r}') \cdot \hat {\bf s} \right)^2 \label{eqn:H1pp}
\end{align}
with $\alpha_s$ defined as in equation \ref{eqn:alpha_mod}.

It is convenient to write time in terms of the precession constant with unitless
$\tau = \alpha_s t$.  The total Hamiltonian including perturbation (using equations
\ref{eqn:hamfirst}, \ref{eqn:ham0}, \ref{eqn:H1pp})
\begin{align}
H(p,\phi,\tau) &= \frac{1}{2} (p-1)^2 - \beta \frac{a'^3}{\Delta^5}  \left( ({\bf r} - {\bf r}') \cdot \hat {\bf s} \right)^2
\label{eqn:Htau}
\end{align}
with unitless coefficient
\begin{equation}
\beta \equiv \frac{m_p}{M_*} \left(\frac{a_s}{a'} \right)^3 \label{eqn:beta}
\end{equation}
primarily dependent on the  ratio $m_p/M_*$ of perturbing planet and stellar masses.

\subsection{Evaluating the perturbation term in the Hamiltonian to first order in inclination}
\label{subsec:firstorder}

From the Hamiltonian in equation \ref{eqn:Htau} we evaluate the rightmost term
using the low eccentricity and inclination literal expansion method with Laplace coefficients
described in section 6.4 by \cite{M+D}.
%$$ \frac{a'^3}{\Delta^5}  \left( ({\bf r} - {\bf r}') \cdot \hat {\bf s} \right)^2,$$
We begin with radial vector ${\bf r} = (x,y,z)$, in terms of orbital elements 
%(2.112, page 51 M+D)
\begin{align}
{\bf r} 
&= 
r\left( \begin{array}{c}
 \cos \Omega \cos (\omega + f) - \sin \Omega \sin (\omega + f) \cos I  \\ 
 \sin \Omega \cos (\omega + f) + \cos \Omega \sin (\omega + f) \cos I  \\ 
 \sin(\omega + f) \sin I
\end{array}  \right), \label{eqn:xyz}
\end{align}
and likewise for the other mass at ${\bf r}'$ using primed orbital elements. 
As is customary  at low inclination %\cite{M+D}
we let 
\begin{align}
\cos I &\approx 1 - I^2/2 \approx  1 - 2 s^2    \nonumber \\
 \sin I & \approx I \approx  2s .\nonumber 
\end{align}
%To carry out the perturbation expansion,
%we use the low eccentricity and inclination literal expansion method with Laplace coefficients
%described in section 6.4 by \cite{M+D}.
%In the expansion we keep terms that are zero-th order in orbital eccentricity ($e,e'$) 
%and first order in inclination ($s,s'$).

%The product 
%\begin{align}
%\hat {\bf r} \cdot \hat{\bf s}&= \cos (\phi - \Omega ) \cos (\omega + f) \sin \theta \nonumber \\
%& + \sin(\phi -\Omega) \cos (\omega + f)  \cos I  \sin \theta  \nonumber \\
%& + \sin(\phi + f) \sin I \cos \theta.
%\end{align}
%exact.
%To zero-th order in eccentricity (or for $e=0$) we can replace the true anomaly $f$ with 
%the mean anomaly $M$ in the above expression.

%\subsection{Zeroth and first order in inclination}

%Using equations \ref{eqn:square}, \ref{eqn:rrp} and \ref{eqn:D0b} 
To zero-th order in eccentricity and first order in inclination
\begin{align}
 \frac{a'^3 }{\Delta^5} & (({\bf r} - {\bf r'})\cdot \hat {\bf s})^2  \approx \nonumber \\
& \Biggl\{ \frac{\sin^2 \theta}{2 } 
\Bigl[ 1 + \alpha^2     \nonumber \\
& ~~~  +  \alpha^2  \cos(2(\lambda - \phi)) + \cos(2 (\lambda' - \phi))  \nonumber \\
& ~~~ - 2 \alpha \cos (\lambda + \lambda' - 2 \phi) 
           -2 \alpha \cos (\lambda - \lambda')  \Bigr]  \nonumber \\
%& ~~~ \times \left[ \sum_{j=1}^\infty b_{5/2}^j(\alpha) \cos (j(\lambda - \lambda'))  + \frac{1}{2} b_{5/2}^0(\alpha) \right] \nonumber \\
&+ 2  \sin\theta \cos \theta  \times \Bigl[  \nonumber \\
&  ~~~~~~ s\alpha^2\sin(2 \lambda - \Omega - \phi)
 +s \alpha^2 \sin(\phi - \Omega) \nonumber \\
&~~~ + s' \sin(2 \lambda' - \Omega' - \phi)
         +s' \sin(\phi - \Omega') \nonumber \\ 
& ~~~ - s\alpha  \sin(\lambda + \lambda'  - \Omega - \phi) \nonumber \\ 
 & ~~~ - s\alpha  \sin(\lambda - \lambda'  - \Omega + \phi)
\nonumber \\
&~~~ - s'\alpha \sin(\lambda + \lambda'  - \Omega' - \phi)  \nonumber \\ 
&~~~ + s'\alpha  \sin(\lambda -\lambda'  - \Omega' + \phi)  \Bigr] \Biggr\} \nonumber \\ 
& ~~~ \times \frac{1}{2} \sum_{j=-\infty}^\infty b_{5/2}^{(j)} (\alpha) \cos (j (\lambda -\lambda'))   .
\label{eqn:H1zerofirst}
\end{align}
%\bigl, \Bigl, \biggl, or \Biggl 
%
The non-secular terms with arguments that are not multiples of $\lambda - \lambda'$
can be rewritten in terms of a single cosine or sine of orbital elements and $\phi$;
\begin{align} &
\frac{\sin^2 \theta}{8} \left[ 
\cos(j \lambda - (j-2) \lambda' - 2 \phi) 
\left(\alpha^2 b_{5/2}^{(j-2)}  + b_{5/2}^{(j)} - 2 \alpha b_{5/2}^{(j-1)} \right) \right. \nonumber \\
& ~~~ + 
\left.
\cos(j \lambda - (j+2) \lambda' + 2 \phi) 
\left( \alpha^2 b_{5/2}^{(j+2)}  +  b_{5/2}^{(j)} - 2 \alpha b_{5/2}^{(j+1)} \right) \right] \nonumber \\
&+   \frac{\sin\theta \cos \theta}{2}  \times \Bigl[  \nonumber \\
&~~~ ~~~\sin(j \lambda -(j-2) \lambda' - \Omega' - \phi) (b_{5/2}^{(j)} - \alpha b_{5/2}^{(j-1)} ) s' \nonumber \\
&~~~ -\sin(j \lambda -(j+2) \lambda' + \Omega' + \phi) (b_{5/2}^{(j)} - \alpha b_{5/2}^{(j+1)} ) s' \nonumber \\
&~~~ +\sin(j \lambda -(j-2) \lambda' - \Omega - \phi) (\alpha^2 b_{5/2}^{(j-2)} - \alpha b_{5/2}^{(j-1)}) s \nonumber \\
&~~~ -\sin(j \lambda -(j+2) \lambda' + \Omega + \phi) (\alpha^2 b_{5/2}^{(j+2)} - \alpha b_{5/2}^{(j+1)} ) s \Bigr].
\label{eqn:epsj2}
\end{align}

Arguments that are rapidly varying will not strongly perturb the spinning planet
as they effectively average to zero.  Only slowly varying arguments
 give resonantly strong perturbations.
The external body has a slower mean motion than the internal one; $n' < n$ 
recalling that $n = \dot \lambda$.
 The slow arguments for positive $j$ must be  those containing $j\lambda - (j+2)\lambda'$
 and so are associated with second order mean motion resonances.
Retaining only  those three arguments  in equation \ref{eqn:epsj2} for a single positive $j$
and using equations   \ref{eqn:H1zerofirst} and \ref{eqn:epsj2} 
we can write a near resonance Hamiltonian  (equation \ref{eqn:Htau}) as
\begin{align}
H(p,\phi,&\tau)^{j:j+2} =  \frac{1}{2}(p-1)^2  \nonumber \\
& - \beta c_{0}^j  p(2-p)  \cos(j \lambda - (j+2) \lambda' + 2 \phi)  \nonumber \\
& -  (1-p) \sqrt{p(2-p)}  \times  \nonumber \\
&~~~~   \Bigl[ \beta c_{s}^j s  \sin(j \lambda -(j+2) \lambda' + \Omega + \phi)    \nonumber \\
& ~~~~~~~ +\beta c_{s'}^j  s'\sin(j \lambda -(j+2) \lambda' + \Omega' + \phi) \Bigr],
\label{eqn:Hamfull_2}
\end{align}
where we have replaced $\theta$ with $p$
using 
\begin{align}
\sin \theta \cos \theta &=(1-p) \sqrt{p(2-p)} \nonumber \\
\sin^2 \theta &= p(2-p). \label{eqn:ptheta}
\end{align}
The unitless coefficients for $j>0$
\begin{align}
c_{0}^j (\alpha) & \equiv \frac{1}{4} \left( 
 \alpha^2   b_{5/2}^{(j+2)} (\alpha) +  b_{5/2}^{(j)} (\alpha) - 2 \alpha b_{5/2}^{(j+1)}(\alpha)  \right)
 \nonumber \\
c_{s}^j (\alpha) & \equiv  \left(  \alpha b_{5/2}^{(j+1)}(\alpha) - \alpha^2 b_{5/2}^{(j+2)} (\alpha)  \right) 
\nonumber \\
c_{s'}^j (\alpha) & \equiv  \left(  \alpha b_{5/2}^{(j+1)}(\alpha)  -  b_{5/2}^{(j)} (\alpha) \right) .
\label{eqn:c0j}
\end{align}
These coefficients are twice those in equation \ref{eqn:epsj2} because we have taken positive and negative
$j$ terms that give the same argument. 
We recall that time is in units of $\alpha_s$, as defined in equation \ref{eqn:alpha_mod},
and the coefficient $\beta$ depends on the mass ratio $m_p/M_*$ (equation \ref{eqn:beta}).

With a given $j$, to be near resonance  $j n \sim (j+2) n'$ 
or $\alpha \sim \left( \frac{j}{j+2}\right)^\frac{2}{3}$.
To aid in applications we have computed the coefficients, $c_0^j, c_s^j, c_{s'}^j$
for $j=1$ to 6 at near resonant semi-major axis ratios $\alpha$ 
and their values are listed in Table \ref{tab:coef}.

Our Hamiltonian (equation \ref{eqn:Hamfull_2}) contains terms that are first order in 
orbital inclination.
This is to be compared to first order mean motion orbital resonances  that lack first order
terms (in $s$) in an expansion of the disturbing function
 and second order inclination mean motion resonances that by definition are proportional to $s^2$.
Previous calculations of spin perturbations have considered the role of secular
frequencies on planet spin orientation by considering
how the orbit variations affect the torque from the star.
In contrast here we have directly evaluated the torque from a nearby planet.
The direct torque, computed here, is proportional to the mass of the perturbing planet  (see equation \ref{eqn:beta}).
Secular perturbations 
scale with the masses of the planets in the system.  So the sizes of these two types 
of spin-resonances are similar.   We estimate that spin resonances associated with mean motion resonances
are about as strong as secular spin resonances.

We could similarly consider how a nearby planet induces perturbations
on the orbit of our spinning planet and then expand the equation
for the torque from the star taking into account these perturbations.
Variations in an expansion of the Hamiltonian in equation \ref{eqn:ham2} 
due to perturbations on the orbit  give
terms with arguments similar to those computed here from an expansion of the Hamiltonian
in equation \ref{eqn:H1}.
The orbit perturbations  arising from the perturbing planet
depends on the ratio $m_p/M_*$
as does our $\beta$, but here our Hamiltonian perturbation contains both
zeroth and first order terms in $s$.  In contrast near a second order mean motion resonance
orbital perturbations are  second order in $e$ and $s$. 
Because it contains zeroth and first order terms in the expansion,
the torque directly exerted
onto the spinning planet from a nearby planet should  be stronger 
than variations on the torque from the star
caused by orbital perturbations from a perturbing planet.   We have neglected
these orbital perturbations, but future work could take them into account.

In our numerical exploration of Styx we found 
two slowly moving angles, $\phi_{s1}, \phi_{s2}$ (defined in equation \ref{eqn:phi_s})
when there were obliquity variations.
These angles can be recognized as arguments in the Hamiltonian in equation \ref{eqn:Hamfull_2}
with index $j=1$, and identifying $\lambda' = \lambda_{Styx}$ and $\lambda = \lambda_{Charon}$.
Our perturbation computation gives terms with arguments consistent with the form
we guessed from the slow moving angles we had seen in our simulations (see \cite{quillen17}).
Our Hamiltonian model effectively describes the spin-resonance we saw in our numerical  simulations.

\begin{table}
\vbox to120mm{\vfil
\caption{\large  Resonance coefficients \label{tab:coef}}
\begin{tabular}{@{}lllllll}
\hline
Resonance & $j$ & $\alpha$ &  $ c_{0}^j (\alpha)$ & $ c_{s}^j (\alpha)$ & $ c_{s'}^j (\alpha)$\\
3:1 & 1 &  0.481 &  0.765 &  1.782 & -4.844 \\
4:2 & 2 &  0.630 &  1.312 &  6.173 & -11.423 \\
5:3 & 3 &  0.711 &  2.027 & 14.270 & -22.378 \\
6:4 & 4 &  0.763 &  2.904 & 27.179 & -38.793 \\
7:5 & 5 &  0.799 &  3.941 & 45.999 & -61.763 \\
8:6 & 6 &  0.825 &  5.139 & 71.831 & -92.386 \\
\hline
Resonance & $j$ & $\alpha$ &  $ c_{e1}^j (\alpha)$ & $ c_{e'1}^j (\alpha)$ \\
2:1 & 1 &  0.630 & -0.971 & -0.384  \\
3:2 & 2 &  0.763 & -2.862 & -0.179  \\
4:3 & 3 &  0.825 & -6.200 &  0.799  \\
5:4 & 4 &  0.862 & -11.380 &  2.943  \\
6:5 & 5 &  0.886 & -18.794 &  6.647  \\
7:6 & 6 &  0.902 & -28.835 & 12.304  \\
\hline
Resonance & $j$ & $\alpha$ &  $ c_{e3}^j (\alpha)$ & $ c_{e'3}^j (\alpha)$ \\
 %resonance j alpha   ce3      ce3-prime
 4:1 & 1 &  0.397 & -1.238 &  2.997  \\
5:2 & 2 &  0.543 & -3.255 &  5.831  \\
6:3 & 3 &  0.630 & -6.546 & 10.170  \\
7:4 & 4 &  0.689 & -11.425 & 16.304  \\
8:5 & 5 &  0.731 & -18.201 & 24.534  \\
\hline
\end{tabular}
{\\ These are coefficients defined in equations \ref{eqn:c0j},
 \ref{eqn:C3e}, and \ref{eqn:C1e}.
We used series expansions  
for the Laplace coefficients  to compute  them.
}}
\end{table}

\subsection{Perturbation terms to first order in eccentricity}
\label{subsec:first_e}

From the Hamiltonian in equation \ref{eqn:Htau} we evaluate the rightmost term
%$$ \frac{a'^3}{\Delta^5} \qquad {\rm and} \qquad \left( ({\bf r} - {\bf r}') \cdot \hat {\bf s} \right)^2,$$
but keeping terms that are first order in orbital eccentricity
and zeroth-order in inclination.
Again we use  the low eccentricity and inclination literal expansion method with Laplace coefficients
described in section 6.4 by \cite{M+D}.
%Using equations \ref{eqn:square}, \ref{eqn:rrp}, \ref{eqn:D0b},  \ref{eqn:square1},
%\ref{eqn:rrp1},  and  \ref{eqn:D1}  we compute
The first order in eccentricity terms that are  added to equation \ref{eqn:H1zerofirst};
\begin{align}
&\frac{a'^3(({\bf r} - {\bf r'})\cdot \hat {\bf s})^2}{\Delta^5}  \stackrel{+}{\approx}   \frac{\sin^2 \theta}{4} \Biggl\{  \nonumber \\
& 
\quad e \alpha^2 \Bigl[ \cos(3 \lambda - \varpi - 2 \phi) \nonumber \\
& \qquad  - 3 \cos(\lambda + \varpi - 2 \phi)
- 2 \cos(\lambda - \varpi)  \Bigr]  \nonumber \\
& - e \alpha
\Bigl[ \cos (2 \lambda + \lambda' - \varpi - 2 \phi)
        +  \cos(2\lambda - \lambda' -  \varpi) \nonumber \\
&  \qquad - 3 \cos (\lambda' + \varpi - 2 \phi) - 3  \cos (\lambda'-\varpi)  \Bigr] \nonumber \\
& -  e' \alpha
\Bigl[ \cos (2 \lambda' + \lambda - \varpi' - 2 \phi)
 +  \cos(2\lambda' - \lambda -  \varpi') \nonumber \\
& \qquad  - 3 \cos (\lambda + \varpi' - 2 \phi) - 3  \cos (\lambda-\varpi') \Bigr] \nonumber \\
& + e'
\Bigl[ \cos(3 \lambda' - \varpi' - 2 \phi) \nonumber \\
& \qquad - 3 \cos(\lambda' + \varpi' - 2 \phi)
- 2 \cos(\lambda' - \varpi')  \Bigr]   \nonumber \\
& \times   \sum_{j=-\infty}^\infty b_{5/2}^{(j)} (\alpha) \cos (j \lambda -j\lambda')   \nonumber  
\end{align}
\begin{align}
& + \Biggl( 1+ \alpha^2+ \alpha^2 \cos(2\lambda -2\phi)
  +  \cos(2\lambda' -2\phi) \nonumber \\
& - 2 \alpha  \cos(\lambda + \lambda'-2\phi) - 2 \alpha \cos(\lambda - \lambda')  \Biggr) \nonumber  \\
& \times  \sum_{j=-\infty}^\infty \Biggl[   \nonumber \\
& ~~~~ e \cos[(j+1) \lambda - j \lambda'- \varpi] \left(- \frac{\alpha}{2} D_\alpha + j \right) b_{5/2}^{(j)}(\alpha) \nonumber \\
& + e\cos[(j-1) \lambda - j \lambda'+ \varpi] \left( -\frac{\alpha}{2} D_\alpha - j \right) b_{5/2}^{(j)}(\alpha) + \nonumber \\
&  + e'   \cos[j \lambda + (1-j) \lambda'- \varpi'] \left( \frac{\alpha}{2} D_\alpha + \frac{5}{2}- j  \right) b_{5/2}^{(j)}(\alpha)  \nonumber \\
 & +  e' \cos[j \lambda - (1+j) \lambda'+ \varpi'] \left( \frac{\alpha}{2} D_\alpha + \frac{5}{2}+ j \right) b_{5/2}^{(j)}(\alpha) \Biggr] \Biggr\}  
\end{align}
Combining arguments and taking only arguments that contain $\phi$ these terms can be written
%\begin{strip}
\begin{align}
\cos[ j \lambda\! - \! (j-3) \lambda'\! - \! \varpi\! - \! 2 \phi]  \;   & \frac{ \sin^2 \theta}{8}
         e \Bigl[ \alpha^2\! \left ( - \frac{\alpha}{2} D_\alpha + j - 2 \right)\! b_{5/2}^{(j-3)}  +
                \alpha \left( \alpha D_\alpha - 2 j + 3 \right)\! b_{5/2}^{(j-2)}  %\nonumber \\
         +   \left( -\frac{\alpha}{2} D_\alpha +  j -1  \right)\! b_{5/2}^{(j-1)} \Bigr] + \nonumber \\
\cos[ j \lambda \! -\! (j+3) \lambda'  \!+\! \varpi \!+ \! 2 \phi]   \;   &  \frac{ \sin^2 \theta}{8}
e \Bigl[ \alpha^2\!  \left ( - \frac{\alpha}{2} D_\alpha - j - 2 \right)\! b_{5/2}^{(j+3)}  +
                \alpha \left( \alpha D_\alpha + 2 j + 3 \right)\! b_{5/2}^{(j+2)}  %\nonumber \\
         +   \left( -\frac{\alpha}{2} D_\alpha -  j -1  \right)\! b_{5/2}^{(j+1)} \Bigr] + \nonumber \\
\cos[ j \lambda \! -\!  (j-3) \lambda' \! -\! \varpi'\! - \! 2 \phi]  \;   &  \frac{ \sin^2 \theta}{8}
e' \Bigl[ \alpha^2\!  \left (  \frac{\alpha}{2} D_\alpha - j  + \frac{9}{2}  \right)\! b_{5/2}^{(j-2)}  +
                \alpha \left( -\alpha D_\alpha + 2 j - 8 \right)\! b_{5/2}^{(j-1)}  %\nonumber \\
         +   \left( \frac{\alpha}{2} D_\alpha -  j  + \frac{7}{2}  \right)\! b_{5/2}^{(j)} \Bigr] + \nonumber \\
\cos[ j \lambda \! - \! (j+3) \lambda' \! + \! \varpi' \!+ \! 2 \phi]  \;  & \frac{ \sin^2 \theta}{8}
e' \Bigl[ \alpha^2\!  \left (  \frac{\alpha}{2} D_\alpha + j  + \frac{9}{2}  \right)\! b_{5/2}^{(j+2)}  +
                \alpha \left( -\alpha D_\alpha - 2 j - 8 \right)\! b_{5/2}^{(j+1)}  %\nonumber \\
        +   \left( \frac{\alpha}{2} D_\alpha + j  + \frac{7}{2}  \right)\! b_{5/2}^{(j)} \Bigr]  + \nonumber \\
\cos[ j \lambda \! -\! (j-1) \lambda' \! + \! \varpi \!- \! 2 \phi]   \;   &  \frac{ \sin^2 \theta}{8}
e \Bigl[   \alpha^2 \!  \left( -\frac{\alpha}{2} D_\alpha - j  - 2 \right)\! b_{5/2}^{(j-1)} 
         +       \alpha \left( \alpha D_\alpha + 2 j + 3 \right)\! b_{5/2}^{(j)} % \nonumber \\
         +  \left ( - \frac{\alpha}{2} D_\alpha - j - 1   \right)\!  b_{5/2}^{(j+1)}  
         \Bigr]   + \nonumber \\
 \cos[ j \lambda  \! - \! (j+1) \lambda' \! -\! \varpi \!+\! 2 \phi]  \;   &  \frac{ \sin^2 \theta}{8}
e \Bigl[  \alpha^2 \!  \left ( - \frac{\alpha}{2} D_\alpha + j - 2   \right)\! b_{5/2}^{(j+1)}  +
                \alpha \left( \alpha D_\alpha - 2 j + 3 \right)\! b_{5/2}^{(j)}  %\nonumber \\
         +  \left( -\frac{\alpha}{2} D_\alpha +  j  - 1 \right)\! b_{5/2}^{(j-1)} \Bigr]  + \nonumber \\
\cos[ j \lambda \! - \! (j-1) \lambda' \! + \! \varpi' \! - \! 2 \phi]   \;  &    \frac{ \sin^2 \theta}{8}
e' \Bigl[  \alpha^2  \! \left (  \frac{\alpha}{2} D_\alpha + j + \frac{1}{2}   \right)\! b_{5/2}^{(j-2)}  +
                \alpha \left( -\alpha D_\alpha - 2 j  \right)\! b_{5/2}^{(j-1)}  %\nonumber \\
         +  \left( \frac{\alpha}{2} D_\alpha +  j  - \frac{1}{2} \right)\! b_{5/2}^{(j)} \Bigr]  +  \nonumber \\
\cos[ j \lambda \! -\! (j+1) \lambda' \! - \! \varpi' \! +\! 2 \phi]  \;  &   \frac{ \sin^2 \theta}{8}
e' \Bigl[  \alpha^2  \! \left (  \frac{\alpha}{2} D_\alpha - j + \frac{1}{2}   \right)\! b_{5/2}^{(j+2)}  +
                \alpha \left( -\alpha D_\alpha + 2 j  \right)\! b_{5/2}^{(j+1)}  %\nonumber \\
         +  \left( \frac{\alpha}{2} D_\alpha -  j  - \frac{1}{2} \right)\! b_{5/2}^{(j)} \Bigr] . %+ \nonumber \\
         \label{eqn:H1e}
\end{align}
%\end{strip}

Inspection of the arguments in equation \ref{eqn:H1e} implies that resonant terms will be important
(with slowly varying arguments)
near first order mean motion resonances, where $j \lambda - (j+1)\lambda'$ is slowly varying 
(with  positive $j$)
and near third order mean motion resonances, where $j \lambda - (j+3)\lambda'$) is slowly varying.
Near a third order mean motion resonance (and to first order in eccentricities and inclinations) 
we can consider a near resonance Hamiltonian 
(similar to equation \ref{eqn:Hamfull_2})  
\begin{align}
H(p,\phi,\tau&)^{j:j+3} \approx  \frac{1}{2}(p-1)^2   \nonumber \\
& -\beta c_{e3}^j e p(2-p)\cos[ j \lambda - (j+3) \lambda' + \varpi + 2 \phi]   \nonumber \\
& -\beta c_{e'3}^j e' p(2-p) \cos[ j \lambda - (j+3) \lambda' + \varpi' + 2 \phi]  . \label{eqn:Hamfull_3}
\end{align}
Near a third order mean motion resonance, we can neglect the terms we previously
computed (zero-th order in $e,e'$ and first order in $s,s'$) because they are only important near 
second order mean motion resonances.
 However, at higher order in $e,e',s,s'$  additional
terms will contribute near all of these mean motion resonances. 

Near a third order mean motion resonance, equation \ref{eqn:Hamfull_3} shows that
the torque directly exerted by the planet
is first order in eccentricity.
The perturbing planet should cause orbital perturbations depending upon the third order
of the eccentricity.   So the variations in the torque from the star due to the orbit variations
are likely to be smaller than the those caused directly from the torque of the perturbing planet.

Near a first order mean motion resonance
\begin{align}
H(p,\phi,&\tau)^{j:j+1} \approx \frac{1}{2}(p-1)^2   \nonumber \\
& -\beta c_{0}^{2j}   p(2-p) \cos [ 2(j\lambda - (j+1) \lambda') + 2 \phi]  \nonumber \\
& -\beta c_{s}^{2j} s  (1-p) \sqrt{p(2-p)} \sin [ 2(j\lambda - (j+1) \lambda') +  \Omega + \phi]  \nonumber \\
&- \beta c_{s'}^{2j} s'  (1-p) \sqrt{p(2-p)} \sin [ 2(j\lambda - (j+1) \lambda') +  \Omega' + \phi]  \nonumber \\
&-\beta c_{e1}^j e p(2-p)\cos[ j \lambda - (j+1) \lambda' - \varpi + 2 \phi]   \nonumber \\
& -\beta c_{e'1}^j e'  p(2-p) \cos[ j \lambda - (j+1) \lambda' - \varpi' + 2 \phi] . 
 \label{eqn:Hamfull_1}
\end{align}
Here zero-th order and first order in inclination terms contribute but they are indexed by $2j$
rather than $j$. 

Near a first order mean motion resonance, equation \ref{eqn:Hamfull_1} shows that
the torque directly exerted by the planet
is first order in eccentricity.
The perturbing planet would also cause orbital perturbations that are first order in eccentricity.
Variations in the torque from the star due to the orbit variations could be similar in size  
to  those caused directly from the torque of the perturbing planet and these could be computed
in future work.

The coefficients for Hamiltonians in equation \ref{eqn:Hamfull_3} and \ref{eqn:Hamfull_1} 
are twice those listed in equation \ref{eqn:H1e} so as to 
include the contribution from a corresponding negative $j$ term that
gives the same argument;  
\begin{align}
 c_{e3}^j & \equiv   \frac{1}{4}
\Bigl[ \alpha^2 \left ( - \frac{\alpha}{2} D_\alpha - j - 2 \right) b_{5/2}^{(j+3)} +   \nonumber \\
     &   \qquad        \alpha \left( \alpha D_\alpha + 2 j + 3 \right) b_{5/2}^{(j+2)}  %\nonumber \\
         +   \left( -\frac{\alpha}{2} D_\alpha -  j -1  \right) b_{5/2}^{(j+1)} \Bigr]  \nonumber \\
  c_{e'3}^j & \equiv    \frac{1}{4}
\Bigl[ \alpha^2 \left (  \frac{\alpha}{2} D_\alpha + j  + \frac{9}{2}  \right) b_{5/2}^{(j+2)}  +\nonumber \\
   &   \qquad                      \alpha \left( -\alpha D_\alpha - 2 j - 8 \right) b_{5/2}^{(j+1)}  %\nonumber \\
        +   \left( \frac{\alpha}{2} D_\alpha + j  + \frac{7}{2}  \right) b_{5/2}^{(j)} \Bigr]  
         \label{eqn:C3e}
\end{align}
and 
\begin{align}
c_{e1}^j & \equiv    \frac{ 1}{4}
         \Bigl[  \alpha^2 \left ( - \frac{\alpha}{2} D_\alpha + j - 2   \right) b_{5/2}^{(j+1)}  + \nonumber \\
&    \qquad               \alpha \left( \alpha D_\alpha - 2 j + 3 \right) b_{5/2}^{(j)}  %\nonumber \\
         +  \left( -\frac{\alpha}{2} D_\alpha +  j  - 1 \right) b_{5/2}^{(j-1)} \Bigr]   \nonumber \\
c_{e'1}^j & \equiv    \frac{ 1}{4}
 \Bigl[  \alpha^2 \left (  \frac{\alpha}{2} D_\alpha - j + \frac{1}{2}   \right) b_{5/2}^{(j+2)}  + \nonumber \\
&  \qquad              \alpha \left( -\alpha D_\alpha + 2 j  \right) b_{5/2}^{(j+1)}  %\nonumber \\
         +  \left( \frac{\alpha}{2} D_\alpha -  j  - \frac{1}{2} \right) b_{5/2}^{(j)} \Bigr]. 
         \label{eqn:C1e}
\end{align}
For low $j$ we computed these coefficients and list them in Table \ref{tab:coef}.

Near a second order mean motion resonance ($j:j+2$) with $j$ an odd integer,
the first order in eccentricity terms do not contribute.  
Consequently equation \ref{eqn:Hamfull_2} remains accurate to first order in eccentricity
for odd $j$ second order  mean motion resonances.

\section{Drifting Toy Hamiltonian models }
\label{sec:spin_toy}

\begin{figure}
\centering
   \includegraphics[trim = 3.0in 2.5in 2.5in 1.4in, clip, width=2.2in]{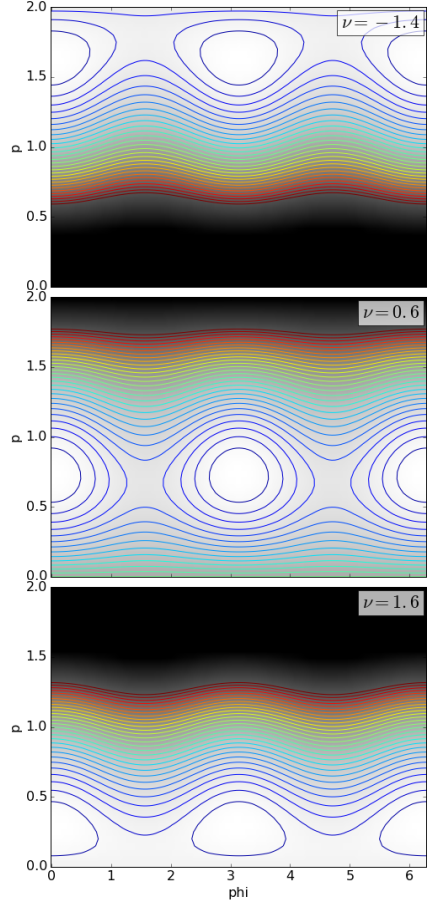}
\caption{Level contours of the Hamiltonian in equation \ref{eqn:K0} 
with a perturbation proportional to $\sin^2 \theta$ with resonance strength $\epsilon=\beta c_{0}^j = 0.05$.
Each subplot shows level contours of the Hamiltonian with a different value of distance to 
resonance $\nu_{2j}$
and with  the value of $\nu_{2j}$ labelled on the upper right. 
For  $|\nu_{2j}| < 2(1 + 2| \epsilon|)$  there are two stable  fixed points at $p$ value that
increases with decreasing $\nu_{2j}$.
\label{fig:K0}}
\end{figure}
% fig1
 
\begin{figure}
\centering
\includegraphics[trim = 2.0in 3.7in 1.2in 1.6in, clip, width=3.7in]{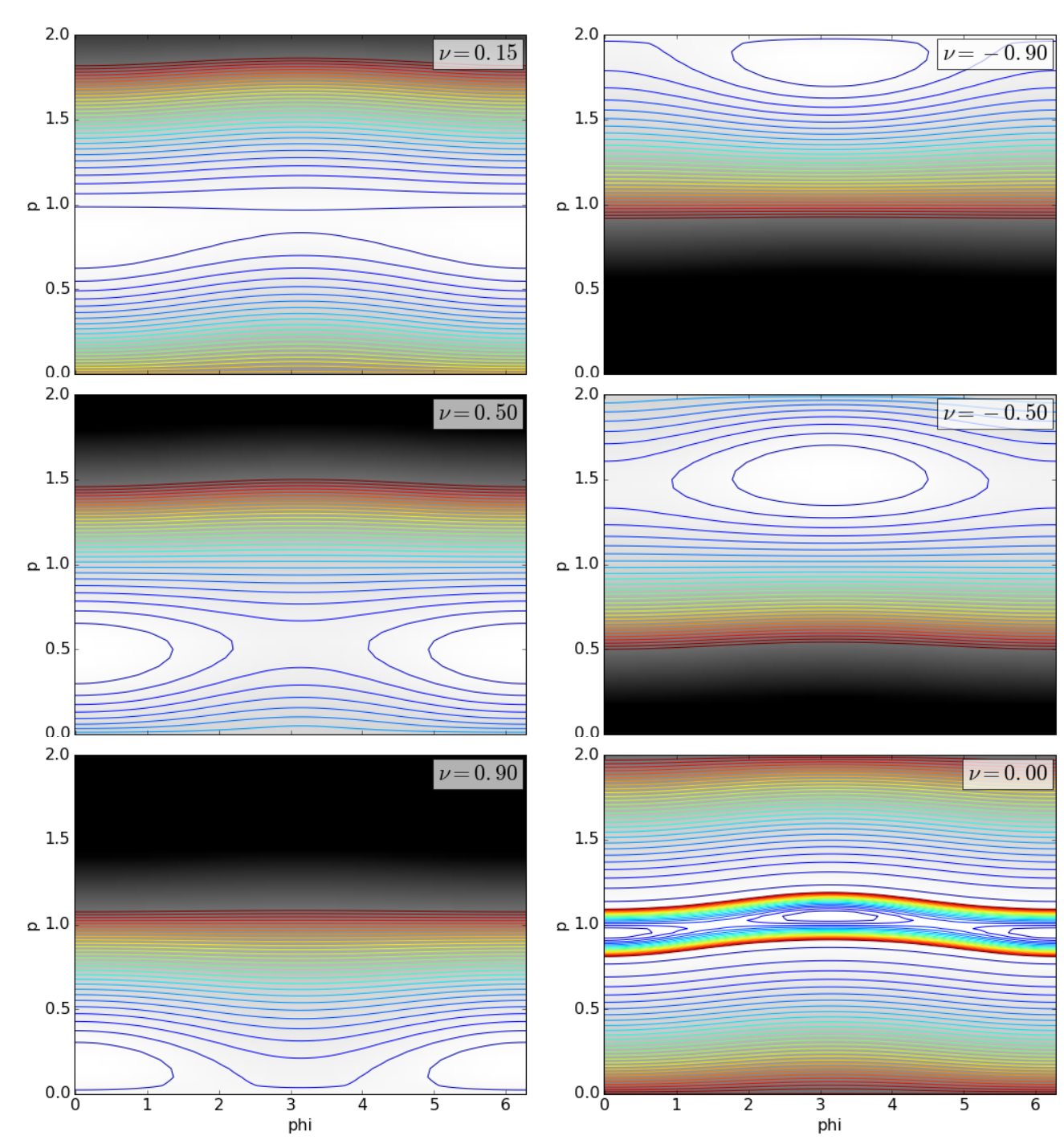}
\caption{Level contours of the Hamiltonian in equation \ref{eqn:K1} 
with a term first order in inclination proportional to $\sin \theta \cos \theta$, 
 with resonance strength $\epsilon  = \beta c_{s}^j s = 0.05$
and for different values of distance to resonance $\nu_{2jo}$.
This Figure is similar to Figure \ref{fig:K0}.
For $\nu_{2jo} \gtrsim $ there is a  stable fixed point at $\varphi=0$.
This one disappears just below $\nu_{2jo}=0$.
For $\nu_{2jo} \lesssim 0$ there is a stable fixed point at $\varphi=\pi$.
This one disappears just above $\nu_{2jo}=0$.
\label{fig:K1}}
\end{figure}
% fig2

The perturbations to the Hamiltonian arising from terms that are zero-th order in inclination and 
eccentricity are proportional to $\sin^2 \theta$ 
(see equation \ref{eqn:epsj2}).
These would be important near first and second order mean motion resonances, 
as shown in the Hamiltonians
 in equations \ref{eqn:Hamfull_2} and  \ref{eqn:Hamfull_1}.
Perturbations that are first order in eccentricity are also  proportional to $\sin^2 \theta$
(see equation \ref{eqn:H1e}) and these would be important near first and third order
mean motion resonances (as in the Hamiltonians
given in equation \ref{eqn:Hamfull_3} and \ref{eqn:Hamfull_1}).
In contrast perturbations that are first order in inclination are proportional to $\sin \theta \cos \theta$
(see equation \ref{eqn:epsj2}).   These are relevant for first and second order mean motion resonances
(as in the Hamiltonians
given in equations \ref{eqn:Hamfull_2} and \ref{eqn:Hamfull_1}).
We have two types of perturbations, those proportional to $\sin^2\theta = p(2-p)$
and those proportional to $\sin \theta \cos \theta = (1-p) \sqrt{p(2-p)}$.
 %equation \ref{eqn:ptheta}
The level curves of these two types of Hamiltonians have different morphologies.

\subsection{Hamiltonian model for a perturbation proportional to $\sin^2\theta$}
\label{subsec:toy0}

Taking the Hamiltonian in equation  \ref{eqn:Hamfull_2}, appropriate for a second order mean motion
resonance we define 
a frequency
\begin{equation} 
\nu_{2j} \equiv \alpha_s^{-1} (j n - (j+2) n').  \label{eqn:nu0}
\end{equation}
Here the spin precession frequency $\alpha_s$ makes frequency $\nu_{2j}$ unitless.
Retaining a single $j$ term zero-th order (in inclination) perturbation term,  the Hamiltonian
in equation \ref{eqn:Hamfull_2} becomes
\begin{align}
H(p,\phi,\tau)_{c0j} &=  \frac{1}{2}(p-1)^2  \nonumber \\
& - \beta c_{0}^j  p(2-p)  \cos(\nu_{2j} \tau + 2 \phi + c)   \label{eqn:Ham:toy0}
\end{align}
with $c$ a constant phase.

A similar Hamiltonian would be derived near a first order mean motion resonance
and retaining only a single first order in eccentricity term using the Hamiltonian
in equation  \ref{eqn:Hamfull_1}. 
In this case the relevant frequency (replacing $\nu_{2j}$) would be one of the following
$\nu = \alpha_s^{-1}(2jn - 2(j+1)n' + \dot \Omega)$, 
$ \alpha_s^{-1}(2jn - 2(j+1)n' + \dot \Omega')$,
$ \alpha_s^{-1}(jn - (j+1)n' + \dot \varpi)$ or 
$ \alpha_s^{-1}(jn - (j+1)n' + \dot \varpi')$
depending upon the argument.
A  Hamiltonian similar to equation \ref{eqn:Ham:toy0} 
can also be derived  near a third order mean motion resonance 
retaining only a single first order in eccentricity term and using the Hamiltonian
in equation  \ref{eqn:Hamfull_3}.   In this case the relevant frequency would be  
$\nu = \alpha_s^{-1}(jn - (j+3)n' + \varpi)$ or $ \alpha_s^{-1}(jn - (j+3)n' + \varpi')$.

Performing a canonical transformation with time dependent 
generating function that is a function of old coordinates and a new momentum
\begin{equation}
F_2(\phi,p',\tau) = \frac{1}{2} (\nu_{2j} \tau + 2 \phi +c)p',
\end{equation}
giving new momentum $p'=p$ equal to the old one and a new angle 
\begin{equation}
\varphi = \frac{1}{2}(\nu_{2j} \tau + 2 \phi + c).
\end{equation}
Transforming the Hamiltonian in equation \ref{eqn:Ham:toy0} we find a new
Hamiltonian in these new coordinates
\begin{align}
K(p,\varphi,\tau)_{c0j} &=  \frac{1}{2}(p-1)^2  + \frac{\nu_{2j} p}{2} - \epsilon  p(2-p)  \cos(2\varphi) 
\label{eqn:K0}
\end{align}
with 
\begin{equation}
\epsilon \equiv \beta c_{0}^j .
\end{equation}
The Hamiltonian is   time independent as long as $\nu_{2j}$ is fixed, 
and $\nu_{2j}$ sets the distance to the center of resonance.
%We have neglected the phase $c$ as it can be removed via shifting $\varphi$ or $\tau$.
 
We can write the frequency (equation \ref{eqn:nu0})
\begin{equation}
 \nu_{2j} \frac{ \alpha_s}{n} = j - (j+2) \left(\frac{a}{a'}\right)^{3/2}. \label{eqn:nu_a}
\end{equation}
During an epoch of  planet migration the  semi-major axes $a,a'$ may drift. 
When two planets  approach each other either $a$ increases or
$a'$ decreases so  the frequency $\nu_{2j}$ decreases.
If the planet orbits separate, $\nu_{2j}$ increases.
%For resonance capture at low obliquity, either
%planet (or both) could migrate, as long as they approach.
 
For various values of $\nu_{2j}$
level curves for the Hamiltonian in equation \ref{eqn:K0} 
are shown in Figure \ref{fig:K0}.  
Fixed points are located at $\varphi=0,\pi$ and $\varphi=\pm \pi/2$.
With $\epsilon>0$, stable fixed points are at $\varphi=0,\pi$ whereas
with $\epsilon<0$ they are at $\varphi=\pm \pi/2$.  
The stable ones are located at  
 $p = [1 + 2 |\epsilon| - \nu/2]/(1+2|\epsilon|)$, and
there are no  fixed points for $|\nu_{2j}| > 2(1 + 2 |\epsilon|)$.
The  fixed point $p$ 
value increases with decreasing frequency $\nu_{2j}$.   
At small $\epsilon$,
the $p$ value for the fixed points range from near 0 to near 2
corresponding to obliquity ranging from near 0 to near $180^\circ$.

\begin{figure}
\includegraphics[width=3.5in]{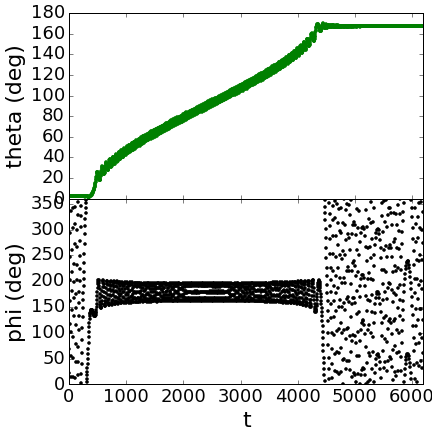}
\caption{Integration of the Hamiltonian in equation
\ref{eqn:K0} with a zero-th order perturbation (in inclination) 
$\propto \sin^2 \theta$ and
resonance strength $\epsilon = 0.01$.  
Here the resonance drifts  with $\dot \nu_{2j} = -0.001$ 
corresponding to the planets slowly migrating closer together. 
Initial conditions are  $(\varphi, p) = (0.4, 0.001)$, and $\nu_{2j} = 2.4$.
The top panel shows obliquity as a function of time whereas
the bottom panel shows the precession angle.
The integrated body is captured into spin resonance at low obliquity and exits
near an obliquity of $180^\circ$.
\label{fig:drift_ham0}}
\end{figure}

We can mimic planet or satellite migration by allowing the frequency specifying distance to resonance
$\nu_{2j}$ to slowly vary.  
We let $\nu_{2j}$ be linearly dependent on time.   We note that $\alpha_s$, setting
our unit of time and the strength of the coefficients, also depends on the semi-major 
axes and spin rate. For the moment we regard them as constants and allow only $\nu_{2j}$ to vary.
As $\nu_{2j}$ decreases, corresponding to the planets migrating so that they approach one another, 
a planet initially at low obliquity could be captured into a stable fixed point 
and lifted in obliquity. Using Hamilton's equations for Hamiltonian of equation \ref{eqn:K0}
and $\epsilon  = 0.01$,
 we integrated  a planet or satellite, within initial low $p=0.001$ 
 (corresponding to an obliquity of $\theta=2.6^\circ$) and with 
 $\dot \nu_{2j} = -0.001$.  The time evolution of $\theta, \varphi$ are shown in Figure \ref{fig:drift_ham0}.
 The planet is captured into a resonant region near a fixed point and lifted to an obliquity of near $180^\circ$
 and then it escapes resonance.  The resonance strength we used is small.
Even when
the resonance is weak and narrow, a planet or satellite could be captured into it and have its obliquity lifted
to high values
as the two planets approach each other and the resonance frequency drifts.

To lift a planet's obliquity from near 0 to $180^\circ$, 
the frequency $\nu_{2j}$ must drop from approximately $ 2$ to approximately $ -2$.    
Using the definition for $\nu_{2j}$ in 
equation \ref{eqn:nu0}, the 
frequency $\nu_{2j} \alpha_s = j n - (j+2)n'$ should drop from $2\alpha_s$ to $-2\alpha_s$
 to lift an initially low obliquity
planet to high obliquity.  If the precession frequency is slow because the planet is nearly
spherical, or is spinning rapidly and has a compact satellite system, 
then the required extent of planetary migration would be small.

In this section we have assumed that resonance strength is time independent, however
 if the spinning planet migrates or spins down, $\alpha_s$ is a function of time
and we have neglected this variation here.  
 The ratio of semi-major axes
$\alpha$ is also a function of time during migration. 
These variations could be taken into account in applications of our toy model
by allowing the resonance strength to be time dependent.
 
\subsection{Hamiltonian model for a perturbation proportional to $\sin\theta \cos \theta$}
\label{subsec:toy1}

We now consider a Hamiltonian model with a single perturbation proportional to $\sin\theta \cos \theta$.
We retain a single $j$  term first order in inclination in the Hamiltonian \ref{eqn:Hamfull_2}.
Using a frequency
\begin{equation} 
\nu_{2jo} \equiv \alpha_s^{-1} ( j n - (j+2) n' + \Omega) 
\end{equation}
we have a simplified Hamiltonian
\begin{align}
H(p,\phi,\tau)_{csj} &=  \frac{1}{2}(p-1)^2  \nonumber \\
& - \beta c_{s}^j  s (1-p)\sqrt{p(2-p)}  \sin(\nu_{2jo} \tau+ \phi + c)   \label{eqn:Ham:toy1}
\end{align}
with constant $c$.
Using  a canonical coordinate transformation with generating function
\begin{equation}
F_2(\phi,p',\tau) =  (\nu_{2jo} \tau +  \phi +c )p',
\end{equation}
we derive a new angle
\begin{equation}
\varphi = \nu_{2jo} \tau + \phi + c
\end{equation}
and a new Hamiltonian
\begin{align}
K(p,\varphi,\tau)_{csj} &=  \frac{1}{2}(p-1)^2 + \nu_{2jo} p  
-  \epsilon (1-p) \sqrt{p(2-p)}    \cos \varphi  \label{eqn:K1}
\end{align}
with resonance strength
\begin{equation}
\epsilon \equiv \beta c_{s}^j s. 
\end{equation}
%Again we have neglected the constant $c$ as we can shift the initial phase. 
The Hamiltonian would look the same if a first order term proportional to $s'$
were used with frequency $\nu= \alpha_s^{-1}(j n - (j+2) n' + \Omega')$ 
instead of $\alpha_s^{-1}(j n - (j+2) n' + \Omega)$,
and with perturbation strength, $\epsilon$, 
equal to $\beta c_{s'}^j s'$ instead of $\beta c_{s}^j s$.
A similar Hamiltonian would be derived near a first order mean motion resonance and
using a single $j$ term that is first order in inclination from the Hamiltonian in equation \ref{eqn:Hamfull_1}.

Level curves for the Hamiltonian in equation \ref{eqn:K1} 
are shown in Figure \ref{fig:K1} for different values of $\nu_{2jo}$ and for $\epsilon = 0.05$.
Fixed points satisfy
\begin{equation}
\nu_{2jo} = -q \pm \epsilon \frac{(1- 2q^2)}{\sqrt{1-q^2}}
\end{equation}
with $q \equiv 1-p$ and the sign of $\epsilon$ is positive for the fixed points at $\phi=0$
 and negative for those at $\phi=\pi$.
For $\nu_{2jo} \lesssim 0$ a stable fixed point is at $\varphi=0$ and has $p<1$.
However just above $\nu_{2jo} = 0$ this fixed point disappears.  For 
$\nu_{2jo} \gtrsim 0 $ there is again a  stable fixed point but at $p>1$.  This one
disappears just below $\nu_{2jo} = 0$.
Near $\nu_{2jo}=0$, two stable fixed points are present at  $\phi=0,\pi$ and the resonant
islands are small.

\begin{figure}
\includegraphics[width=3.5in]{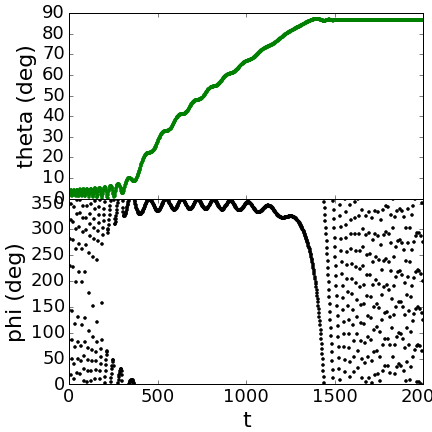}
\caption{Integration of the Hamiltonian in equation
\ref{eqn:K1} with a first order perturbation (in inclination) proportional
to $\sin \theta \cos \theta$ and resonance strength
$\epsilon = 0.01$.  
Here the resonance drifts  with $\dot \nu_{2jo} = -0.001 $ for planets slowly migrating closer together.
Initial conditions are  $(\varphi, p) = (1.5, 0.001)$, and $\nu_{2jo} = 1.4$.
The top panel shows obliquity as a function of time whereas
the bottom panel shows the precession angle.
The integrated body is captured into spin resonance at low obliquity and exits
near an obliquity of $90^\circ$.
\label{fig:drift_ham1}}
\end{figure}

As before we integrate the equations of motion for a slowly drifting system.
We plot a planet trajectory in Figure \ref{fig:drift_ham1} using the Hamiltonian in equation
\ref{eqn:K1},  $\epsilon = 0.01$, $\dot \nu_{2jo} = -0.001$, 
initial conditions  $(\varphi, p) = (1.5, 0.001)$ and $\nu_{2jo} = 1.4$.
Again we find that resonance capture is possible for a particle initially
at low obliquity.  However because the fixed point
disappears near an obliquity of $90^\circ$ the planet must escape resonance
near this value rather than near $180^\circ$ as for the  perturbation that is 
proportional to $\sin^2 \theta$ (explored in subsection \ref{subsec:toy0}). 

In this section we have assumed that resonance strength is time independent.
However variation in semi-major axis ratio, $\alpha$, precession rate $\alpha_s$
and orbital inclination can affect the resonance strength.

%fig3

%fig4

\subsection{Adiabatic Limits for Resonance Capture}

When the drift from migration is too fast or not adiabatic,
a particle in orbit will jump across mean motion resonance
rather than capture into resonance \cite{quillen06}.    
 At drift rates below a critical drift rate, the resonance
can capture at high probability.  This critical drift rate is approximately equal to the square of the  
resonance libration frequency  \cite{quillen06}. 
There is  a critical initial eccentricity, below which capture into mean motion
resonance is assured when drift is adiabatic \cite{borderies84}.   
The drift rate defining the adiabatic limit and the limiting eccentricity ensuring capture
can be estimated from
the Hamiltonian by considering the dimensions of the coefficients \cite{quillen06}.
By considering the form
of the Hamiltonians in equations \ref{eqn:K0}, \ref{eqn:K1} at low $p$ we can similarly estimate
the drift rate required for capture into spin resonance.

At low $p$ (low obliquity) the perturbation term in the Hamiltonian of equation \ref{eqn:K0} 
(with perturbation $\propto \sin^2 \theta$)
is proportional to $\epsilon 2 p \cos 2 \phi$, so the drifting resonance behaves like a second order
mean motion resonance that is proportional to $e^2$ or the Poincar\'e momentum variable $\Gamma$. 
The resonance libration frequency 
at low $p$ is $\omega_{lib} \propto | \epsilon|$.
In analogy
to the Hamiltonian for mean motion resonance, we expect that capture into resonance
for the drifting Hamiltonian of equation \ref{eqn:K0}
is likely for drift rates $|\dot \nu_{2j}| \lesssim \epsilon^2$ (the square of the libration frequency, and
following  \cite{quillen06,mustill11})
and for initial momentum
$p \lesssim |\epsilon|$  (the critical momentum below which capture is assured for adiabatic drift; 
see \cite{borderies84,quillen06}).

In contrast the resonant term for the Hamiltonian in \ref{eqn:K1}, 
with perturbation $\propto \sin \theta \cos \theta$,
at low $p$
is proportional to $\sqrt{p}$ and so this resembles a first order mean motion resonance,
proportional to $e$ or the square root of the Poincar\'e momentum variable, $\sqrt{\Gamma}$. 
The libration frequency $\omega_{lib} \propto |\epsilon|^\frac{2}{3}$ (following \cite{quillen06,mustill11}).
The resonance should capture at a drift rate $|\dot \nu_{2jo}| \lesssim |\epsilon|^\frac{4}{3}$
(the square of the libration frequency)
and for initial $p \lesssim |\epsilon|^\frac{2}{3}$.
 
\begin{figure}
\centering
   \includegraphics[width=3.5in]{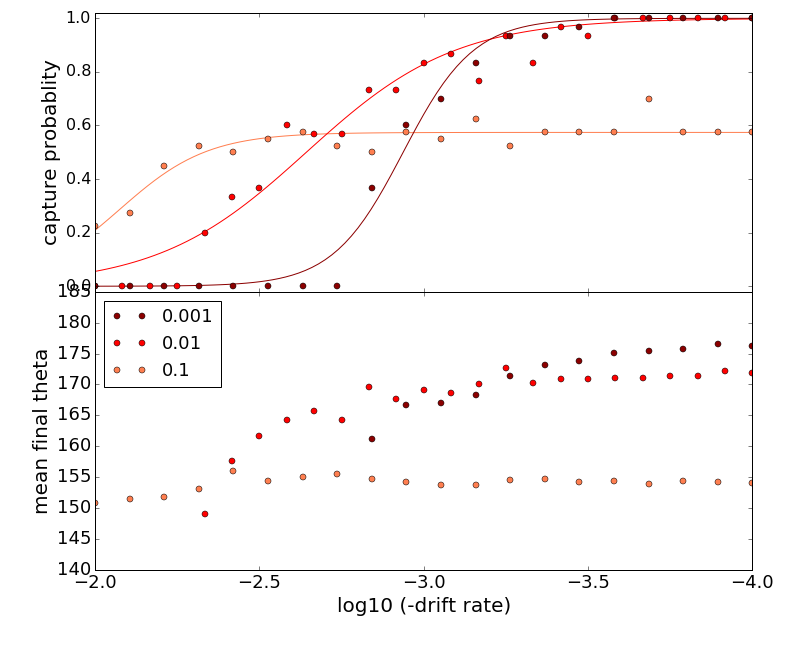}
  \includegraphics[width=3.5in]{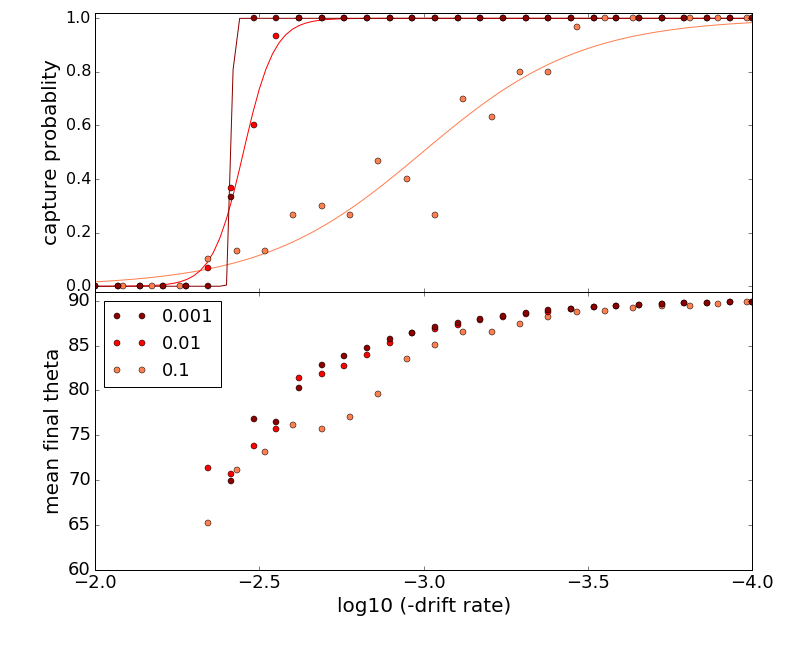}
\caption{a) The top panel shows capture probabilities for
the Hamiltonian in equation \ref{eqn:K0} 
with resonant term $\propto \sin^2 \theta$ with resonance strength $\epsilon = 0.01$ as a function of
drift rate for three different initial obliquities or $p$ values.
The initial $p$ values 0.001, 0.01, 0.1, shown in the key on the lower left,
 correspond to initial obliquities of
2.6, 8.1 and 26$^\circ$. 
  The bottom panel
shows the average final obliquities (in degrees) when capture took place for the same initial
$p$ values.  
 The $x$ axes show the log$_{10}$ of the absolute
value of the drift rate $|\dot \nu_{2j}|$.  
The drift rate is higher on the left hand side and gives lower capture probabilities.
Each point is computed from an average of 30 integrations.
b) Similar to a) except for the Hamiltonian in equation \ref{eqn:K1}, with 
resonant term $\propto \sin\theta \cos \theta$ and with
$x$ axis showing $\log_{10} |\dot \nu_{2jo}|$.
\label{fig:cap}}
\end{figure}

To estimate coefficients for the limits we show capture probabilities for a range of drift rates
(different $\dot \nu_{2j}, \dot \nu_{2jo}$ values)
for the Hamiltonians  in  equations \ref{eqn:K0} and \ref{eqn:K1} in Figure \ref{fig:cap}.
In both cases we set perturbation strength $\epsilon = 0.01$.   
We computed the capture probabilities for three different initial $p$ values, 0.001, 0.01, and 0.1
corresponding to initial obliquities of 2.6, 8.1 and $26^\circ$.
Each point shown in Figure \ref{fig:cap} was computed from an average of
30 integrations where each integration was begun outside
of resonance at a randomly chosen initial $\phi$, chosen from a uniform
distribution spanning $[0,2\pi]$. The scatter likely arises from dependence
on phase when the drifting system reaches resonance \cite{quillen06,mustill11}.
To reduce dependence on phase and reduce
scatter in these plots we also chose random initial  $\nu_{2j}, \nu_{2jo}$ values,
ensuring that we began outside of resonance and from a uniform distribution
with a width of 0.5.   For those integrations that captured into spin resonance
(as determined by a large increase in obliquity or $p$)
we computed the mean final obliquity (of the integrations that captured into resonance)  
and these are shown as points on the bottom
panels in Figure \ref{fig:cap}.
The solid lines show fits of a hyperbolic tangent function to the capture probability
and are used to determine where the transition from low to high probability
takes place.

Figure \ref{fig:cap} shows that for low initial $p$
capture takes place at $|\dot \nu_{2j}| \sim 10^{-3}$ for the $\sin^2 \theta$ resonance
with  Hamiltonian in equation \ref{eqn:K0}.  Using the dependence on 
$\epsilon^2$ we estimate that
\begin{equation}
\left|\dot \nu_{2j} \right|_{\sin^2 \theta} \lesssim 10 \epsilon^2  \label{eqn:driftlim0}
\end{equation}
for capture into the $\sin^2 \theta$ resonance.

Capture takes place at
$|\dot\nu_{2jo} |\sim   4 \times 10^{-3}$ for the $\sin\theta \cos \theta$
resonance (Hamiltonian of equation \ref{eqn:K1})
giving
\begin{equation}
|\dot \nu_{2jo} | _{\sin \theta \cos \theta} \lesssim 2 |\epsilon|^\frac{4}{3}  \label{eqn:driftlim1}
\end{equation}
for capture into the $\sin \theta \cos \theta$ resonance.

The $\sin \theta \cos \theta$ resonance (Figure \ref{fig:cap}b) exhibits a sharper sensitivity to drift
rate than the $\sin^2 \theta $ one (Figure \ref{fig:cap}a), 
consistent with previous studies showing that first order
mean motion resonances have a more abrupt transition in capture probability near the adiabatic limit
\cite{quillen06}.

Following resonance capture, 
Figure \ref{fig:cap} shows that after escaping resonance the final obliquity 
is somewhat sensitive to initial obliquity and drift rate. 
For the $\sin^2 \theta$ resonance, higher initial obliquity gave
lower final obliquity after resonance capture.
For both types of resonances, nearer the adiabatic limit and
at higher drift rates, the final obliquities are somewhat lower.

How long does it take to lift the obliquity in one of these spin resonances?
For the $\sin^2 \theta$ resonance, once captured into resonance, $p$ must drift
from 0 to 2. Within a factor of a few  the total time in resonance is $ 1/|\dot \nu_{2j}|$.
For the $\sin \theta \cos \theta$ resonance the total time in resonance is equivalent.
Restoring units, once captured into resonance at low obliquity, the time to lift 
 the obliquity to near $180^\circ$ for the $\sin^2 \theta$ resonance 
 or near $90^\circ$  for the $\sin \theta \cos \theta $  resonance is
\begin{equation}
t_{lift} \sim \frac{1}{|\dot \nu | \alpha_s}. \label{eqn:tlift}
\end{equation}
where $\nu$ is the appropriate frequency (either $\dot \nu_{2j}$ or $\dot \nu_{2jo}$).

We consider how the distance to spin resonance, $\nu$, might vary when two
planets are in mean motion resonance and migrating.
Once captured into mean motion resonance, eccentricity or and inclination can continue
to increase as the system migrates.   While in resonance, a resonant argument 
librates about a constant value.  For example,
an inclination 3:1 mean motion resonance contains the argument
$ \lambda - 3 \lambda' + 2 \Omega$.
With this argument librating the frequency $\omega = n - 3 n' + 2 \dot \Omega$ approximately
averages to zero.
However, the frequencies setting distance to the spin resonance near the second-order
mean motion resonance, $\nu_{2j}$ or $\nu_{2jo}$, are not the same
as $\omega$.   The difference between   $\nu_{2j}$ and $\omega$ is $2 \dot \Omega$ 
for the spin resonance $\propto \sin^2 \theta$ and the difference between $\nu_{2jo}$ and 
$\omega$ is 
 $\dot \Omega$ for the spin resonance $\propto \sin \theta \cos \theta$.
The precession rate $\dot \Omega$ is sensitive to inclination and would vary
if the orbital inclination increases within mean motion resonance.  The same is true
for $\dot \varpi$ as eccentricity increases within mean motion resonance.
As the orbital inclination or eccentricity increases within mean motion resonance, 
the frequencies  setting the distance to spin resonances  would vary.  
As spin  resonance frequencies can vary for a  
body evolving within a mean motion resonance, spin resonances can be encountered,
causing either spin resonance capture
or a jump in obliquity as the spin resonance is crossed.
 
%fig5

\section{Application to Pluto and Charon's minor satellites}
\label{sec:pc}

We consider spinning minor satellites Styx and Nix, that have orbits
exterior to Charon. With an exterior spinning body,
the unitless coefficient $\beta$  (defined in equation \ref{eqn:beta})
is approximately equal to the mass ratio
of Charon and Pluto, $\beta \sim 0.1$.
Styx is near the 3:1 mean motion resonance with Charon and 
Nix near the 4:1 mean motion resonance with Charon.

We first consider Styx, with an orbital period of 20 days. 
As Styx is near the 3:1 second order mean motion resonance we consider the
Hamiltonian in equation \ref{eqn:Hamfull_2} with $j=1$.
Inspection of Table \ref{tab:coef}
shows that there is a zero-th order (in inclination and eccentricity)
perturbation with  coefficient
$c_0^1(\alpha) = 0.76$ giving a resonance term
$\propto \sin^2 \theta$.   There are two terms first order in 
orbital inclination with larger coefficients, $c_s^1 = 1.782$
and $c_{s'}^1 = -4.84$ and giving  resonance terms $\propto \sin \theta \cos \theta$.  
The $c_0$ term need only be multiplied by $\beta$ giving resonance strength
$\epsilon = 0.1 \times 0.76 = 0.08$ for the Hamiltonian in equation \ref{eqn:K0}.
The $c_s^1$ and $c_{s'}^1$ coefficients 
should be multiplied by the orbital
inclinations of Styx and Charon, respectively, to estimate the strength
of the $\sin \theta \cos \theta$ resonance.  Working in a coordinate
frame aligned with Styx's orbit (and with the obliquity measured
with respect to Styx's orbit normal), we use the inclination of Charon
and need only the $c_{s'}^1$ coefficient.
Taking into account $\beta=0.1$, we estimate a resonance coefficient
strength $\epsilon = 0.1 \times c_{s'}^1 = 0.48 I_{Charon}$
where $I_{Charon}$ is the inclination of Charon's orbit relative
to Styx's.  With an inclination of $I_{Charon} = 6^\circ = 0.1$ radians
we find $\epsilon = 0.05$.  The resonant  perturbation term is $\propto \sin \theta \cos \theta$
and the relevant Hamiltonian is equation \ref{eqn:K1}.

The two resonance strengths, $\propto \sin^2 \theta$ and that 
$\propto \sin \theta \cos \theta$, have similar strengths
when Charon's inclination is  $\sim 8^\circ$.
With a moderate inclination, the $c_{s'}^1$ resonance
might be more important than the $c_0^1$ one.
After capture,  
Styx should exit the $\sin \theta \cos \theta$ resonance with
an obliquity near $90^\circ$, as we saw in our simulations (section 5 by \cite{quillen17})
 and as illustrated
in Figure \ref{fig:drift_ham1}.
In contrast after capture into the $c_0$ resonance, $\propto \sin^2 \theta$,
Styx should exit the resonance near an obliquity of $180^\circ$.

Which spin resonance is encountered by Styx first as Charon drifts away from Pluto?
The two resonances are not  on top of each other as the $\sin^2\theta$ resonance has
argument with
 frequency  $\nu_{2j} \alpha_s = n_{Styx} - 3 n_{Charon}$, whereas for the $\sin \theta \cos \theta$
resonance the argument frequency is $\nu_{2jo} \alpha_s = n_{Styx} - 3 n_{Charon} - \Omega_{Charon}$.
As $\dot \Omega_{Charon} <0$, for an outward drifting Charon, Styx should
 encounter the $\sin \theta \cos \theta$ resonance first and that may explain why
 the obliquity was lifted to near $90^\circ$ in our simulations rather than $180^\circ$ 
 (see section 5 by \cite{quillen17}).  
 At high obliquity, the $\sin^2 \theta$
 resonance could still be important.   Perhaps this resonance, 
 or evolution involving both perturbation terms could account for the higher
 than $90^\circ$ obliquities of Nix and Hydra measured by \cite{weaver16}.

With resonance strength $\epsilon \sim 0.05$ the drift rate  in the $\sin \theta \cos \theta$
resonance (using equation \ref{eqn:driftlim1})
$|\dot \nu_{2jo}| \lesssim 0.04 $ and giving a constraint on the migration rate
$|\dot n| \lesssim 0.04 \alpha_s^2$ or migration on a timescale
$t_{mig} \gtrsim 30 \left(\frac{n}{\alpha_s}\right)^2 n^{-1}$.
Here either Charon can move away from Pluto or equivalently Styx could migrate inward.
As discussed in section 2.4 of \cite{quillen17},
the precession rate for all the minor satellites is $\alpha_s \sim \frac{n}{2}  \frac{P_s}{P_o}$
where $\frac{P_s}{P_o}$ is the ratio of spin period to orbital period.
Currently $\frac{P_o}{P_s} \sim 6.2, 13.6$ for Styx and Nix, respectively.
This gives a constraint on the migration timescale
\begin{equation}
 t_{mig} = \frac{a}{\dot a} \gtrsim 30 \left( \frac{P_s}{P_o}  \right)^2 P_o \end{equation}
or a few hundred times greater than the orbital period.
This limiting migration rate is not slow and would be  
satisfied during an epoch of circumbinary disk evolution (e.g., \cite{kenyon14}).
The total time required for the lift in obliquity to take place (using equation \ref{eqn:tlift})
would be only of order a hundred orbital periods or a few thousand years,  easily satisfied, and consistent
with the rapid obliquity lifts seen in our simulation.
The extent of migration required is roughly the semi-major axis times the ratio $P_s/P_o$
so is of order 1/6 the semi-major axis, taking Styx's current value for this ratio.
If Charon migrates rather than Styx, the extent of migration required is lower
due to the index 3 on $\lambda_{Charon}$ in the resonant angle and on $n_{Charon}$ in 
the associated frequency $\nu_{2j}$.

Nix is near the third order  4:1 mean motion resonance and the Hamiltonian 
we consider is that in equation \ref{eqn:Hamfull_3}. The terms that are important
are the coefficients that are first order in eccentricity, $c_{e3}^1$
and $c_{e'3}^1$ and with $j=1$.    However this resonance is $\propto \sin^2 \theta$
and our simulation showed resonance escape near an obliquity of $\sim 90^\circ$.
We would attribute this behavior to higher order terms, such as proportional to $es'$,  that
we neglected from our computations in section \ref{subsec:first_e}.   
%However a comparison of terms
%in Table \ref{tab:coef} suggests that the coefficients would be a few times
%larger than the $c_{e3}^1$ or $c_{e'3}^1$ coefficients.   
We have checked that the expansion to first order in $es'$ (or $e's'$) would
give terms proportional to $\sin \theta \cos \theta$ and with argument
similar to a third-order mean motion resonance.  For the 4:1 resonance, dependence
on inclination and eccentricity for the $\sin \theta \cos \theta $ term implies that the spin resonance
strength would be weaker for Nix than Styx and so a slower migration rate and longer
time would be needed to tilt Nix than  Styx.  However, the constraints on these quantities
for Styx were easy to satisfy, so the mechanism for tilting Nix is also likely to be effective.
If these spin resonances operated on Styx and Nix, the near $90^\circ$ obliquities imply that
Charon's orbit was relatively inclined during the migration as the strongest perturbation
terms $\propto \sin \theta \cos \theta$
are first order in orbital inclination.

Our toy model considers each resonant term separately, but likely both $\sin^2 \theta$
and $\sin \theta \cos \theta$ terms
are present and simultaneously important for these spin resonances.  
We have not yet explored higher inclinations or slower
migration rates in our mass-spring model simulations, leaving this  for future work.  
At moderate inclination both Styx and Nix
 might initially be captured
into a $\sin \theta \cos \theta$ resonance but escape or could be subsequently pushed
higher by a $\sin^2 \theta$ resonance, possibly accounting for
Nix's $123^\circ$ \cite{weaver16} and higher than $90^\circ$ obliquity.

Kerberos is near a 5:1 mean motion resonance with Charon and Hydra near 
a 6:1 mean motion resonance.  
Similar spin resonances could have operated on both of these satellites.
Our Hamiltonian models in sections \ref{subsec:firstorder} and \ref{subsec:first_e}
imply that the spin resonances near the 5:1 and 6:1 mean motion resonances
would be higher than first order in inclination and eccentricity.
They would be weaker, require  slower migration rates for capture
and longer time to evolve within resonance to lift the obliquities.
Our previous numerical simulations did not show Kerberos tilted  by the same spin resonance
mechanism as Styx and Nix, but perhaps this is 
because we had not run longer simulations at  slower migration rates (required for resonance
capture because the spin resonances would be weaker).
Since we only ran simulations of three bodies (Pluto, Charon and a satellite)
satellite interactions were not present, however these can increase eccentricities or inclinations and the 
associated spin resonance strengths.   

We previously speculated that a mean motion resonance between Nix and Hydra
could affect Hydra's obliquity \cite{quillen17}.   However the strength of such a resonance
depends on the mass ratio of Nix and Pluto and this would make the 
spin resonances  4 to 5 orders of magnitude
weaker than those involving Charon and so very weak.  The resonance strengths computed
here arise from the direct torque applied from a orbiting point mass, in this setting
from Nix directly onto Hydra.  Indirectly Nix could excite
the orbit of Hydra due to a 3:2 first order mean motion resonance between them
and the torque from Pluto and Charon arising from the orbit perturbations of Hydra
might affect Hydra's spin.   This  perturbation would probably be weak as it also depends
on the mass ratio of Nix and Pluto, however if important it
 would be most effective near a first order mean motion resonance such as the 3:2 
  mean motion resonance (first order 
in eccentricity and between Nix and Hydra), as discussed at the end of section \ref{subsec:first_e}).
More likely is that interactions between the satellites increased the orbital eccentricities and inclinations
making the spin resonance induced by Charon stronger.

\section{Application to Uranus}
\label{sec:uranus}

The  successful {\it Nice} model \cite{tsiganis05} and its variants 
(e.g., \cite{morbi09,nesvorny11,nesvorny12,nesvorny15,deienno17}), postulate an
epoch or epochs of planetary migration beginning with planets near
or in mean motion resonance \cite{morbi07}.   During planet migration, secular resonances
and encounters between planets can alter planet spin orientation or obliquity \cite{ward04}.  
Likewise the current
 obliquities of the giant planets could give clues about the extent and speed
 of  planetary migration  during  early epochs  of Solar system evolution (e.g., \cite{boue10,brasser15}).
 
%Saturn has spin precession rate approximately equal to the vertical secular eigenfrequency 
%associated with Neptune. 
%\cite{ward04,hamilton04} proposed an elegant scenario, 
%in which slow evolution within secular spin-orbit resonance is responsible for tilting Saturn 
%to its current $27^\circ$ obliquity. 
%%see \cite{helled09}for a potential problem with this scenario.
%Jupiter's spin precession rate is close to the vertical secular eigenfrequency associated with Uranus,
%\cite{ward06} but its low $3^\circ$ obliquity means that its obliquity evolution could not be similar to Saturn's.
%The constraint that Saturn is currently tilted but not Jupiter is satisfied by a small 
%fraction of the potential planetary migration scenarios associated
%with the {\it Nice} model \cite{brasser15}. 

%Uranus, with its odd obliquity of 98$^\circ$, has long presented a puzzle.

The large obliquity of Uranus  has primarily been attributed
to  a  tangential or grazing  collision  with  an Earth-sized   proto-planet at  the
end of the epoch of accretion 
(e.g., \cite{safronov69,korycansky90,slattery92,lee07,parisi08,parisi11}).
However, \cite{boue10} proposed a collisionless scenario involving
an additional, but now absent, satellite.   Proposed is that
a close encounter at the end of the era of migration ejected this satellite 
 while Uranus was at high orbital inclination, and following orbital inclination damping
 the  planet was left at high obliquity.  
Drift of secular spin resonances, explaining  Saturn's obliquity, might
also account for Uranus's obliquity \cite{rogo16}.
%\cite{ward04,hamilton04}

We consider the possibility that during an early epoch of planetary migration
Uranus was   captured into a mean motion resonance with another giant planet 
and while in it, its obliquity was lifted due to spin resonance.
This scenario would give an alternative non-collisional scenario accounting
for Uranus's high obliquity.
The spin resonance strengths (see equation \ref{eqn:beta})
depend on the mass ratio of perturbing
planet to that of the central star 
and this is at most  the ratio of Jupiter's mass to that of the Sun or $\sim 10^{-3}$. 
For Uranus external to its perturber, our parameter $\beta$ is equal to this
mass ratio.   
Using coefficients in Table \ref{tab:coef} we estimate a 5:3 resonance strength
$\epsilon = \beta c_0^{j=3} \sim 2 \times 10^{-3}$, a 4:3 resonance strength
$\epsilon = \beta c_{e1}^{j=3} e_p \sim 6 \times 10^{-3} e_p$ with $e_p$ the eccentricity of
the inner perturbing planet.
At most these coefficients are of order $10^{-2}$ and so 
the spin resonance strengths
are likely to be about 10 times weaker than we considered in the Pluto-Charon system.
 Equations \ref{eqn:driftlim0} or \ref{eqn:driftlim1} and \ref{eqn:tlift} imply that the time required to lift Uranus
would be of order 1000 times its precession period.  

Taking into account its satellite system, Uranus's precession period is currently
approximately $10^8$ years \cite{ward75}.   The precession period is so slow that
there is not enough time during a {\it Nice} model epoch of planetary migration
for the spin resonances considered here to tilt the planet.
If Uranus had a much heavier and more extended satellite system \cite{mosqueira03a,mosqueira03b} then
its precession period would be reduced.  In this case we could consider
capture into a 3:2, 4:3 or 5:3 resonance with Saturn and a spin resonance
associated with one of these mean motion resonances.  However the
resonance strengths depend on eccentricity and  inclination, making them
even weaker.  The time required to 
lift the planet remains long, of order the age of the Solar system, and requiring
a timescale longer than the postulated era of migration. 
 Uranus and Saturn migrating in such proximity
are unlikely to remain stable but the spin resonance would require hundreds
of spin precession periods to tilt Uranus.   We conclude that the type of spin
resonance  explored here
cannot account for Uranus's high obliquity.

The spin secular resonance mechanism explored by \cite{ward04}
involves drift of secular spin resonance.   The secular spin resonances
are about the same size as the mean-motion/spin precession resonances.   
But Uranus currently is near a secular
resonance, that associated with the vertical secular eigenfrequency 
associated with Neptune.   The spin-secular resonance could have operated on a timescale
of billions of years.  For a mean-motion/spin precession resonance to operate the planet would
need to be near a mean motion resonance for billions of years but Uranus is no longer
near one and was probably not near or in one for as long a time period.

\section{Summary and Discussion}

We have explored a Hamiltonian model for the dynamics of the principal axis of rotation
of an orbiting planet or satellite assuming that the spinning body
remains rapidly spinning about
its principal inertial axis.  We have computed the torque exerted on the spinning body
from a point mass also in orbit about the central mass (a star if the our spinning object
is a planet or a planet if our spinning object is a satellite).  Unlike many previous
studies (e.g., \cite{colombo66}), we do not average over
the orbital period.    We have computed perturbations from the orbiting point mass to first order
in orbital inclinations and eccentricities (but not their product).   
The perturbation terms are either proportional to $\sin^2 \theta$ or to $\sin \theta \cos \theta$
with obliquity $\theta$.
Taking a single Fourier component (of the perturbation) we derive 
the Hamiltonians for the spin orientation shown in equations \ref{eqn:Hamfull_2}, \ref{eqn:Hamfull_3},
and \ref{eqn:Hamfull_1} that are relevant near second order, third order and first order
mean motion resonances, respectively, but affect obliquity and spin precession rate.  
The resonant arguments for the resonance near a
second order mean motion resonance in equation \ref{eqn:Hamfull_2}
are consistent with slowly moving angles (equation \ref{eqn:phi_s}) 
we previously saw in  numerical simulations of Styx 
when its obliquity varied \cite{quillen17}.
Our Hamiltonian model provides a framework for estimating the strengths of spin resonances
involving a mean motion resonance and the precession angle of a spinning body.

Numerical integrations
 of one of these one-dimensional Hamiltonians  (equations \ref{eqn:Hamfull_2}, \ref{eqn:Hamfull_3},
or \ref{eqn:Hamfull_1}) containing a single resonant perturbation term show
that  if the resonance drifts, a spinning body initially at low obliquity can be captured into spin resonance
and its obliquity lifted to near $180^\circ$ or $90^\circ$ depending upon whether
the perturbation term is $\propto \sin^2 \theta$ or $\sin \theta \cos \theta$.
The $\sin \theta \cos \theta$ requires non-zero orbital inclination of the spinning
body with respected to the perturbing one.
We estimate the maximum drift rate allowing spin resonance capture and 
the timescale required to reach the maximum obliquity when the body escapes resonance.
Resonance capture into these spin resonances
only takes place if the perturbing mass and spinning body have
approaching orbits, similar to capture into mean motion orbital resonance.

We applied our Hamiltonian model to a migrating Pluto-Charon satellite system.
The spin resonance
seems capable of accounting for the large obliquity variations we previously saw
in our simulations of the spin evolution of satellites Styx and Nix near the 3:1 and 4:1 mean motion
resonances with Charon.  Outward migration of Charon
or inward migration of Styx and Nix could have let initially low obliquity Styx and Nix
be captured into a spin resonance that lifted their obliquities.   As Styx and Nix
have obliquities near $90^\circ$ and not near $180^\circ$,  the capture would have involved
a resonance proportional to $\sin \theta \cos \theta$ and so requires
Charon's orbit to be inclined by a few degrees with respect to the orbits of Styx and Nix.
Due to Charon's large mass, the resonances are sufficiently strong that the constraints
on the migration rate for resonance capture are loose and the time needed 
to tilt the satellites is short, of order only a thousand years.
Similar spin resonances could operate on Kerberos and Hydra, though they
would be higher order in eccentricity and inclination and so weaker.

We explored whether Uranus could be tilted by a similar mechanism during
an early epoch when Uranus might have been in a first order mean motion resonance with
another giant planet such as Saturn.  However Uranus's spin precession period
is  long enough that tilting the planet would require billions of years.  
Since Uranus
probably did not spend much time in or near mean motion resonance with another planet, 
this type of spin resonance is unlikely to account for Uranus's current high obliquity.  

As the giant planets in the Solar system are not currently in or near strong mean motion 
resonances, the mean-motion/spin precession resonance 
 was justifiably neglected from previous studies.
However the spin-resonances discussed here may be important in migrating rapidly spinning
satellite and planetary systems, such as Pluto and Charon's and possibly migrating compact exoplanet
or satellite
systems prior to tidal spin down.  The approximate Hamiltonian model presented here 
could aid in interpretation of future simulations of spin evolution in such settings.

We explored toy Hamiltonian models containing a single resonant argument.
However real systems are likely to be affected by multiple terms, each dependent on 
an argument that varies at a similar frequency.   Spin evolution could 
be chaotic due to these nearby resonances.  Interaction between
the terms might allow resonance escape at obliquities between 90 and $180^\circ$ possibly
accounting for the  123$^\circ$  obliquity of Nix \cite{weaver16}.
The spin resonances are important
near mean motion resonances, and so future study of spin evolution 
where these spin resonances are important should also
consider the orbital dynamics in or near mean motion resonance (e.g., \cite{voyatzis14}).

\vskip 0.1truein

\begin{acknowledgements}
%We acknowledge helpful discussions with Santiago Loane.
% and Beno\^it Noyelles if they are  not collaborators.
We warmly thank our referee, Gwena\"el Bou\'e, for kindly finding and helping us correct our errors.
Her time is very much appreciated.
We gratefully acknowledge support from the Simons Foundation 
and the hospitality of the
Leibniz Institut f\"ur Astrophysik, Postdam.
BN acknowledges the financial support of the contract Prodex CR90253 from the Belgian Science Policy Office. %(BELSPO).
\end{acknowledgements}

{}

\end{document}